\documentstyle[psfig]{cupconf}

\title[Color--Selected High Redshift Galaxies]{Color--Selected High Redshift Galaxies\\
and the HDF}

\author{Mark Dickinson}

\affiliation{The Johns Hopkins University and STScI}

\def\spose#1{\hbox to 0pt{#1\hss}}

\def\kms{\ifmmode {\rm\,km\,s^{-1}}\else ${\rm\,km\,s^{-1}}$\fi}

\def\kmsmpc{\ifmmode {\rm\,km\,s^{-1}\,Mpc^{-1}}\else ${\rm\,km\,s^{-1}\,Mpc^{-1}}$\fi}

\def\ergps{\ifmmode {\rm\,erg\,s^{-1}}\else ${\rm\,erg\,s^{-1}}$\fi}
\def\ergpscm2{\ifmmode {\rm\,erg\,s^{-1}\,cm^{-2}}\else
    ${\rm\,erg\,s^{-1}\,cm^{-2}}$\fi}

\def\deg{\ifmmode {^{\circ}}\else {$^\circ$}\fi}
\def\degr{\ifmmode {^{\circ}}\else {$^\circ$}\fi}
\def\degs{\ifmmode {^{\circ}}\else {$^\circ$}\fi}

\def\h3Mpc{h^{-3}{\rm Mpc}^3}

\def\arcsec{\ifmmode {^{\prime\prime}}\else $^{\prime\prime}$\fi}
\def\asec{\ifmmode {^{\prime\prime}}\else $^{\prime\prime}$\fi}
\def\arcmin{\ifmmode {^{\prime}}\else $^{\prime}$\fi}
\def\amin{\ifmmode {^{\prime}}\else $^{\prime}$\fi}

\def\secper{\ifmmode \rlap.{^{s}}\else $\rlap{.}{^{s}} $\fi}
\def\minper{\ifmmode \rlap.{^{m}}\else $\rlap{.}{^m} $\fi}
\def\secspt{\ifmmode \rlap.{^{\prime\prime}}\else
    $\rlap.{^{\prime\prime}}$\fi}
\def\arcsper{\ifmmode \rlap.{^{\prime\prime}}\else
    $\rlap.{^{\prime\prime}}$\fi}
\def\minspt{\ifmmode \rlap.{^{\prime}}\else
    $\rlap.{^{\prime}}$\fi}
\def\arcmper{\ifmmode \rlap.{^{\prime}}\else
    $\rlap.{^{\prime}}$\fi}
\def\spose#1{\hbox to 0pt{#1\hss}}
\def\simlt{\mathrel{\spose{\lower 3pt\hbox{$\mathchar"218$}}
     \raise 2.0pt\hbox{$\mathchar"13C$}}}
\def\simgt{\mathrel{\spose{\lower 3pt\hbox{$\mathchar"218$}}
     \raise 2.0pt\hbox{$\mathchar"13E$}}}

\def\refindent{\par\noindent\parskip=2pt\hangindent=3pc\hangafter=1 }

\def\aj{{AJ}}
\def\apj{{ApJ}}

\def\mnras{{MNRAS}}

\def\ref#1;#2;#3;#4 {\refindent{#1,} {#2}, #3, #4}

\def\book#1;#2;#3 {\refindent{#1, }{in {\it{#2},} }{#3}}
\def\reference#1{\refindent{#1}}

\def\U300{\ifmmode{U_{300}}\else{$U_{300}$}\fi}
\def\B450{\ifmmode{B_{450}}\else{$B_{450}$}\fi}
\def\V606{\ifmmode{V_{606}}\else{$V_{606}$}\fi}
\def\I814{\ifmmode{I_{814}}\else{$I_{814}$}\fi}

\begin{document}

\maketitle

\begin{abstract}

The quality, depth, and multi--color nature of the Hubble Deep Field
images makes them an excellent resource for studying galaxies at
$z > 2$ using selection techniques based on the presence of
the 912\AA\ Lyman break.   I present a descriptive review of this 
method and of the properties of the objects which it identifies, 
and summarize spectroscopic progress on galaxies with $2 < z < 4$ 
in the HDF.  Using ground--based and HDF samples of Lyman break 
galaxies I discuss the luminosity function of galaxies at 
$z \approx 3$, and consider the effects of extinction on the
star formation rates that are derived from the UV luminosity
information.  Infrared observations of the HDF provide data 
on the rest--frame optical properties of $z \approx 3$ galaxies, 
which are briefly described.

\end{abstract}

\firstsection % if your document starts with a section,

\section{Introduction}

Although the study of galaxies at high redshift neither begins nor ends
with the Hubble Deep Field (HDF), this conference has demonstrated
the ways in which the HDF has served to focus the attention of the 
community on the properties of galaxies at $z > 2$.  In part this 
is because the HDF imaging data was obtained through 
several filters, permitting the use of color selection techniques to 
isolate and study populations of galaxies at various redshifts.  
% The availability of imaging data taken through 
% four independent optical bandpasses (plus additional follow--up from 
% the ground at other wavelengths) has encouraged investigators to explore 
% various means of correlating galaxy colors with redshift, as well as
% to use multicolor photometric indices to interpret the physical
% properties of distant galaxies.   
Because the HDF images are so
deep, colors can be measured with unusually high precision (and
in a spatially resolved fashion {\it within} individual galaxies) 
for objects which ordinarily would be considered very faint for 
ground--based telescopes.  Additionally, the HDF images easily
detect galaxies at magnitudes well beyond the spectroscopic limits
of even the largest telescopes.  The desire to understand
their nature requires {\it some} idea of their distances, and it is 
therefore tempting to look for photometric means of estimating redshifts 
without the benefit of ordinary spectroscopy.  

For these reasons, many presentations at this symposium 
have considered various applications of ``photometric redshift'' estimation 
in the HDF and in other data sets.  One such method takes advantage of 
the ubiquitous 912\AA\ Lyman limit discontinuity, which is redshifted
into the HDF bandpasses at $z \simgt 2$.  In recent years, color selection 
based on the Lyman limit has developed into a highly successful means 
of detecting galaxies at large redshifts, as I will review below.
In designing the HDF observations, our working group at STScI incorporated 
F300W imaging into the four--filter scheme for two reasons, one scientific 
and one purely practical.  First, such data offers the potential for Lyman 
break selection of high redshift galaxies.   Second, we wished to take 
advantage of the ``bright'' portions of the orbit during Continuous View Zone 
visibility, when scattered earthlight severely impacts WFPC2 imaging at 
redder wavelengths.   The reduced amplitude of the scattered background 
at UV wavelengths, and the fact that F300W imaging with WFPC2 is not 
normally background limited anyway, made bright--time observing through
that bandpass a suitable use for these otherwise disadvantageous 
observing intervals.

I begin with a descriptive review of the Lyman break technique,
illustrated using data from the HDF.   Although the HDF Lyman break
galaxy sample is much smaller than that which has been identified
from ground--based imaging studies (which cover much larger solid
angles), the excellent photometric precision of the HDF 
data and the large percentage of Lyman break galaxies present at the 
fainter magnitude limits which it probes makes it quite useful for 
illustrating the principles of the method.  In \S 2.1 I summarize the 
current status of spectroscopy on HDF Lyman break galaxies.  In \S 3 
I discuss some of the statistical properties of Lyman break galaxies, 
concentrating on their numbers, luminosities, and colors, and inferences 
that can be derived from these measurements.  Finally, \S 4 discusses
the rest--frame optical properties of $z > 2$ galaxies in the HDF
using deep infrared imaging data.  In his contribution to this volume, 
Mauro Giavalisco discusses our ground--based Lyman break galaxy sample 
in greater detail, and addresses the spectroscopic, morphological, 
and clustering properties of Lyman break galaxies, which I will largely
neglect here.

In this paper, HDF galaxy selection and object names are based on 
the catalog of Williams et al.\ (1996).  Optical photometry is reported 
on the AB magnitude system, with the WFPC2 bandpasses indicated by 
\U300, \B450, \V606 and \I814.  In general I have used new, revised 
photometry of the Williams et al.\ catalog objects based on optimized 
apertures which maximize signal--to--noise for color measurement.
This has the advantage of allowing robust Lyman break galaxy selection 
to somewhat fainter magnitudes than was previously possible.

\section{Lyman break selected galaxies in the HDF}

Figure 1 illustrates the principle of the Lyman break color selection 
technique using a galaxy from the HDF as an example.  In the absence of 
dust extinction, an actively star--forming galaxy should have a blue 
continuum at rest--frame ultraviolet wavelengths, nearly flat in $f_\nu$ 
units.  Blueward of the 912\AA\ Lyman limit, however, photoelectric 
absorption by intervening sources of neutral Hydrogen
will sharply truncate the spectrum.   This hydrogen may be located in 
the photospheres of the UV--emitting stars themselves, in the 
interstellar medium of the distant galaxy, or along the intergalactic 
sightline between us and the object.  When observed at some large redshift, 
the rest--frame Lyman limit of a galaxy shifts between some pair of 
bandpasses (e.g. the WFPC2 \U300 and \B450 filters in figure 1), and 
the galaxy ``drops out'' when viewed through the bluer filter because 
of the suppression of its flux.  In addition to the Lyman continuum 
absorption, the cumulative effect of the Lyman~$\alpha$ forest lines 
introduces an additional spectral break shortward of Lyman~$\alpha$ 
at the emission redshift of the galaxy.  This flux suppression is 
increasingly strong at higher redshifts as the forest thickens, and 
introduces its own color effects, particularly for galaxies at $z > 3$.

\begin{figure}
\centerline{\psfig {figure=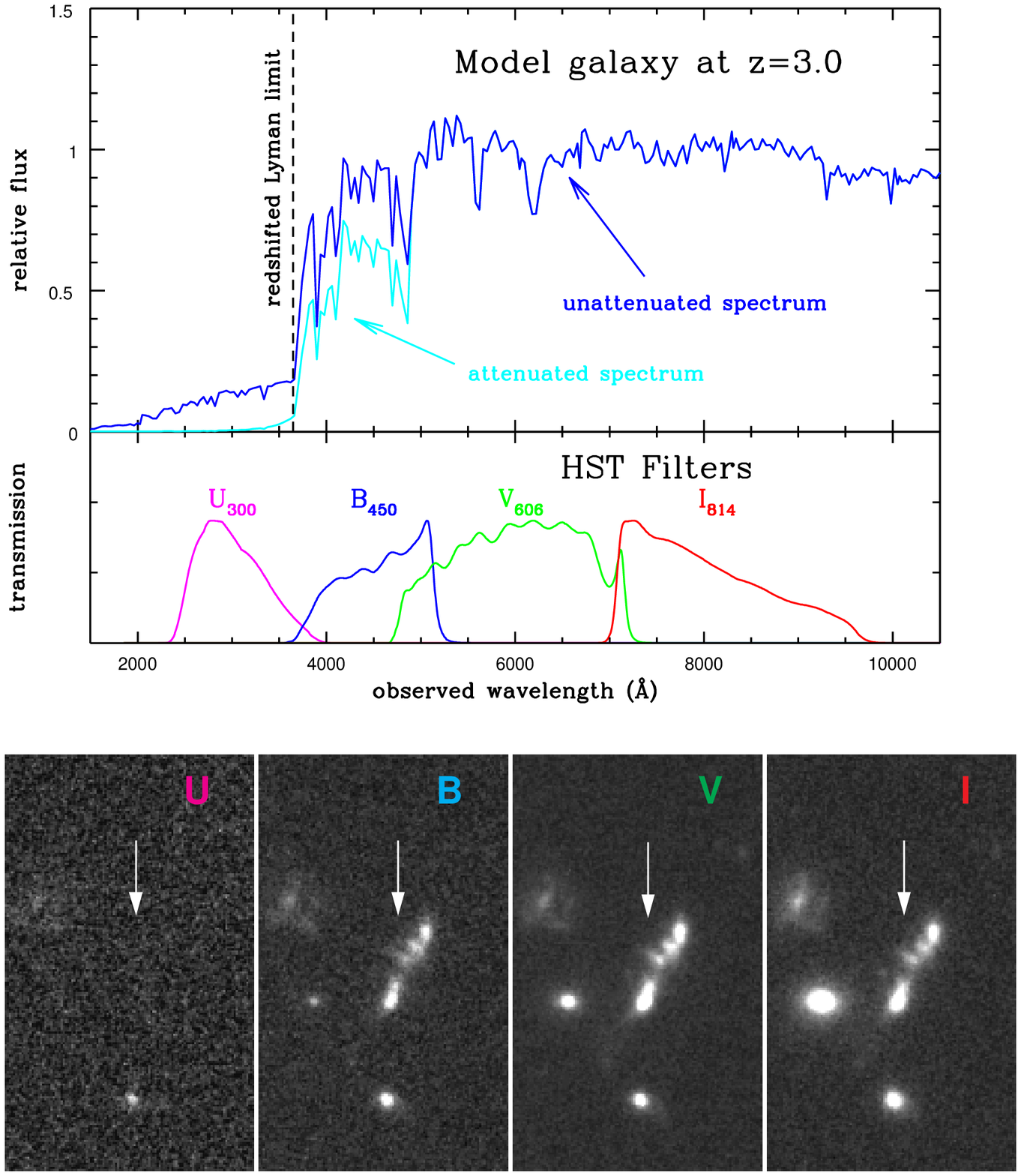,height=6in}}
\caption{Illustration of the Lyman break technique as applied to
the Hubble Deep Field.  The upper panel shows a model spectrum
of a star forming galaxy observed at $z = 3$.  Its flat UV continuum 
is truncated by the 912\AA\ Lyman limit, which is redshifted between 
the \U300 and \B450 filters (WFPC2 bandpasses shown below spectrum).
In addition to photospheric absorption in the UV--emitting stars, 
the effects of intergalactic neutral hydrogen further suppress the 
continuum in the \U300 and \B450 bands.  At bottom, an HDF galaxy is shown 
in the four WFPC2 bandpasses.  Clearly visible at \I814, \V606 and \B450, 
it vanishes in the \U300 image.  This galaxy has been spectroscopically 
confirmed to have $z = 2.8$.}
\end{figure}

Color selection based on the effects of the Lyman limit and Lyman~$\alpha$
forest has been used for many years in surveys for distant 
QSOs (e.g. Warren et al.\ 1987).  The method was applied to the 
study of distant galaxies by Guhathakurta et al.\ (1990) 
and Songaila, Cowie \& Lilly (1990), who used it set limits on 
the number of star--forming galaxies at $z \approx 3$ in faint galaxy 
samples.  Steidel \& Hamilton (1992) and Steidel, Hamilton \& Pettini 
(1995) reported the detection of significant numbers of high redshift 
galaxy candidates using this method.  Spectroscopic confirmation 
of their redshifts was first presented by Steidel et al.\ (1996a),
and WFPC2 images of select examples were published by 
Giavalisco et al.\ (1996b).  To date, the majority of 
Lyman break selected galaxies come from the $U_n G {\cal R}$ 
survey of Steidel et al., which has identified more than a thousand 
candidates and has spectroscopically confirmed (as of this writing) 
more than 400 galaxies at $z \approx 3$.  This survey is discussed 
in greater detail by Giavalisco in this volume, and in the present 
paper I will often refer to it as the ``ground--based sample'' 
to distinguish it from HDF--selected objects.   

Although Lyman break galaxies in the HDF are fewer in number than 
those in the ground--based sample, the quality and depth of the HDF 
imaging data offer a number of advantages.  The HDF can be used to 
detect Lyman break galaxies at fainter apparent magnitudes 
than has been achieved in ground--based data, and the precision of 
the \B450, \V606 and \I814 photometry ensures small random
errors on color measurements.   Moreover, the depth and resolution 
of the WFPC2 imaging permits detailed morphological study of these 
objects.   The primary {\it disadvantage} of the HDF is its 
small field of view, and hence the rather small comoving 
volume which it samples.  This limits its utility for statistical
studies (e.g. of luminosity functions, redshift distributions, etc.), 
as small number statistics, galaxy clustering, and field--to--field 
variations may introduce significant uncertainties.

\begin{figure}
\centerline{\psfig {figure=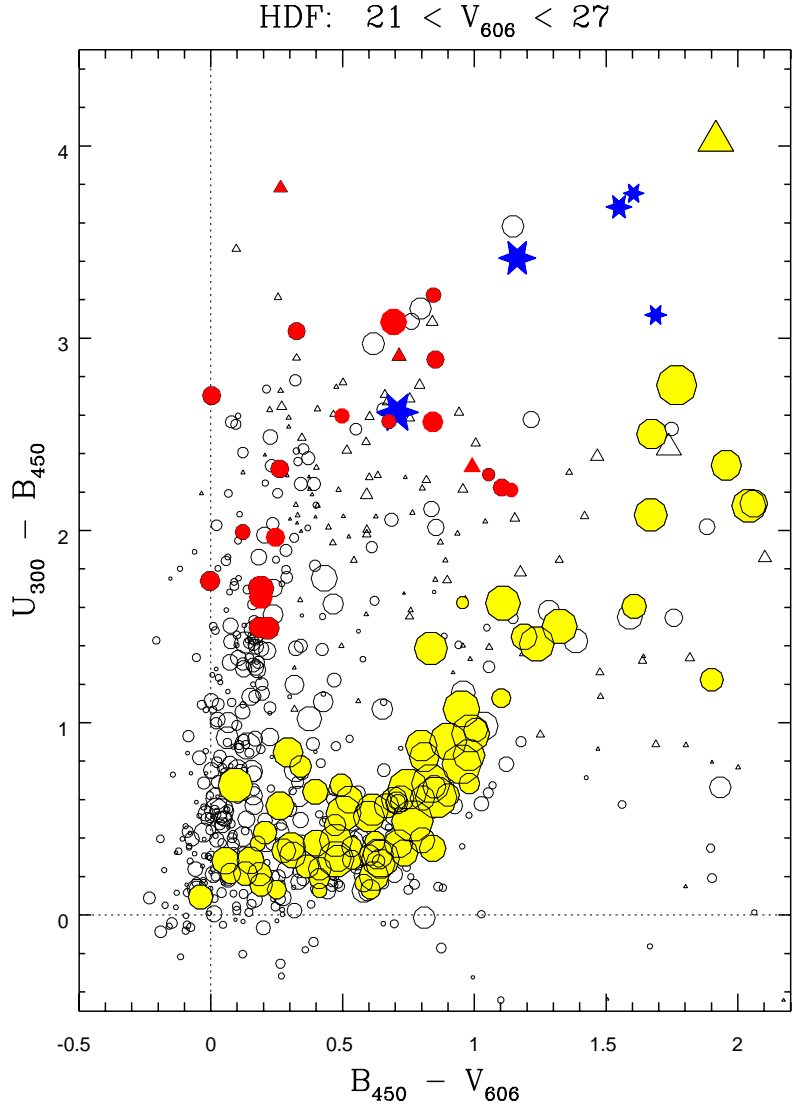,height=6in}}
\caption{Color--color diagram of faint galaxies in the Hubble Deep Field,
illustrating the ``plume'' of Lyman break objects rising from 
$\U300 - \B450 = \B450 - \V606 = 0$.  These are nearly all galaxies at 
$z > 2$.  Spectroscopically confirmed objects in this redshift range are 
shown as darker filled symbols;  galaxies with measured redshifts $z < 2$ 
are shown as light filled circles, and stars are indicated by star--shaped
points.  Triangles mark lower limits ($1\sigma$ to the $\U300 - \B450$ 
color for objects undetected in \U300.  Symbol size scales inversely with 
apparent \V606 magnitude.}
\end{figure}

\subsection{Color selection of Lyman break galaxies in the HDF}

My own favorite ``view'' of the Hubble Deep Field is that shown 
in figure 2, a color--color diagram of galaxies in the HDF.
One of the most prominent features of this diagram is the dramatic
``plume'' of galaxies rising nearly vertically from the zero 
color point (i.e. flat spectrum galaxies) up toward very red 
$\U300 - \B450$ colors.   These are the high redshift, star forming 
galaxies -- objects whose 912\AA\ Lyman discontinuities are entering 
into and passing through the F300W bandpass, shifting them into a portion 
of color--color space which is unpopulated by low redshift objects.

Figure 3 illustrates the redshift dependent effect of the Lyman 
limit and Lyman~$\alpha$ spectral ``breaks'' on the colors 
of galaxies with spectroscopically confirmed redshifts $z > 2$.
To first order, the galaxies in the high redshift ``plume'' form 
a redshift sequence ordered by $\U300 - \B450$ color.  
In practise, variation in individual galaxy spectra shuffle this 
sequence somewhat.   Dynamic range for the \U300 photometry also
affects this ordering, as it is only possible to set limits on
the $\U300 - \B450$ colors of fainter objects, and the numerical
values of these limits therefore depend on the apparent magnitude of 
the galaxies.   The lower panel of figure 3 shows the increasing
effect of the Lyman~$\alpha$ forest (and, at $z \simgt 3.5$, of the
Lyman limit) on the \B450 - \V606 color.  This effect was used by
Lowenthal et al.\ (1997) as an additional criterion for identifying
high redshift galaxy candidates (cf. also Fruchter, this volume).
Using color selection criteria such as those of Steidel et al.\
(1996a,b;  see also below), the application of a color cut
such as $\B450 - \V606 \leq 1.2$ results in an upper redshift
bound $z \simlt 3.5$.  It is this reddening of $\B450 - \V606$ which
is responsible for the ``tilt'' of the plume in the $UBV$ color--color
diagram shown in figure 2 -- the higher redshift \U300 dropout galaxies
``fan out'' toward redder \B450 - \V606 colors.  Figure~4 shows
a different 2--color diagram of HDF galaxies, this time using 
\V606 - \I814 colors which are relatively unaffected by Ly$\alpha$
absorption at $z < 4$.   Here, the ``plume'' remains more or less
vertical.

The Lyman break color technique is a simple form of photometric 
redshift selection.   Here we do not attempt to accurately estimate 
the redshifts of individual objects, but simply to use color criteria
to select galaxies in a particular redshift interval and exclude 
foreground (and background) objects.  The redshift selection function
of the method depends on the particular color criteria adopted,
on the intrinsic dispersion in the ultraviolet spectral properties of 
star forming, high redshift galaxies, on cosmic variance in the
intergalactic transmission along different lines of sight, and
on the distribution of photometric measurement errors.  This redshift
selection function can be estimated using spectral models along with
realistic simulations of photometric errors,  or can be measured directly 
by obtaining enough spectroscopic redshifts to define it empirically.  
We now know that the Lyman break galaxy population exhibits strong 
clustering in redshift space (Steidel et al.\ 1998;  cf. also Giavalisco, 
this volume, and figure 7 below).  Therefore in order to determine the 
redshift selection function empirically one must study many independent 
sightlines in order to average over the effects of large scale structure.  
We have done this for our ground--based survey, and thus feel that we 
understand the redshift selection function quite well.  For the HDF, 
we have only the one ``realization'' of the redshift distribution, and 
thus the empirical selection function cannot properly be 
determined -- here we must rely to some degree on models, although
these models can be informed by the information on galaxy spectral 
properties which has been gained from the ground--based samples.

\begin{figure}
\centerline{\psfig {figure=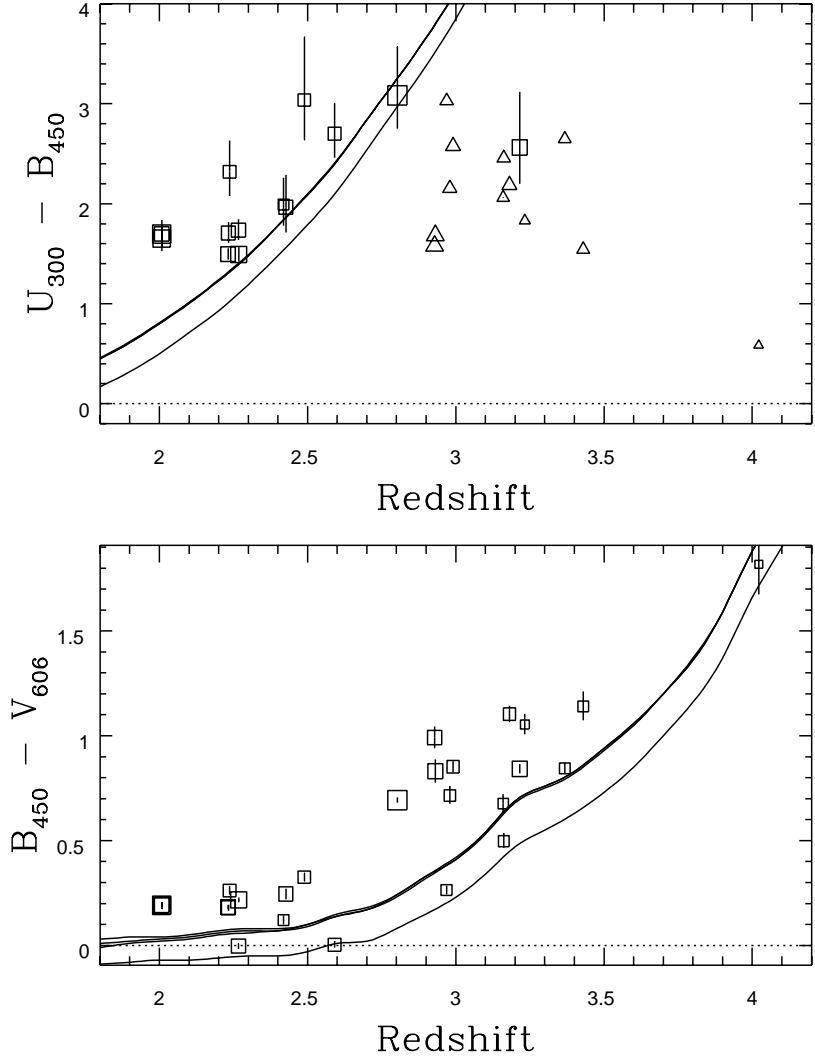,height=6in}}
\caption{\U300 - \B450 and \B450 - \V606 versus redshift for 
spectroscopically confirmed $z > 2$ galaxies in the HDF.  These plots
illustrate the effect of Lyman limit and Lyman~$\alpha$ absorption
on the colors of high redshift galaxies.  The solid lines show 
predicted colors of actively star--forming galaxies of various sorts 
using the Madau (1995) prescription for mean intergalactic transmission.
In the top panel, triangles mark lower color limits for galaxies
with $S/N < 2$ in the \U300 band.  All galaxies with $z < 2.9$ are 
detected in \U300, while all but one galaxy with $z > 2.9$ have
$S/N < 2$.  (The exception, 4-858.0, is formally detected with
$S/N \approx 2.5$;  this may partially result from the red leak
in the \U300 filter or from systematic measurement error, but could 
also indicate photometric contamination from a foreground object.)  
The lower panel shows the progressive reddening of the \B450 - \V606 
color with redshift, primarily due to the effect of the Lyman~$\alpha$
forest.  (Compare with figure 10 for galaxies from the ground--based 
sample.)}
\end{figure}

\begin{figure}
\centerline{\psfig {figure=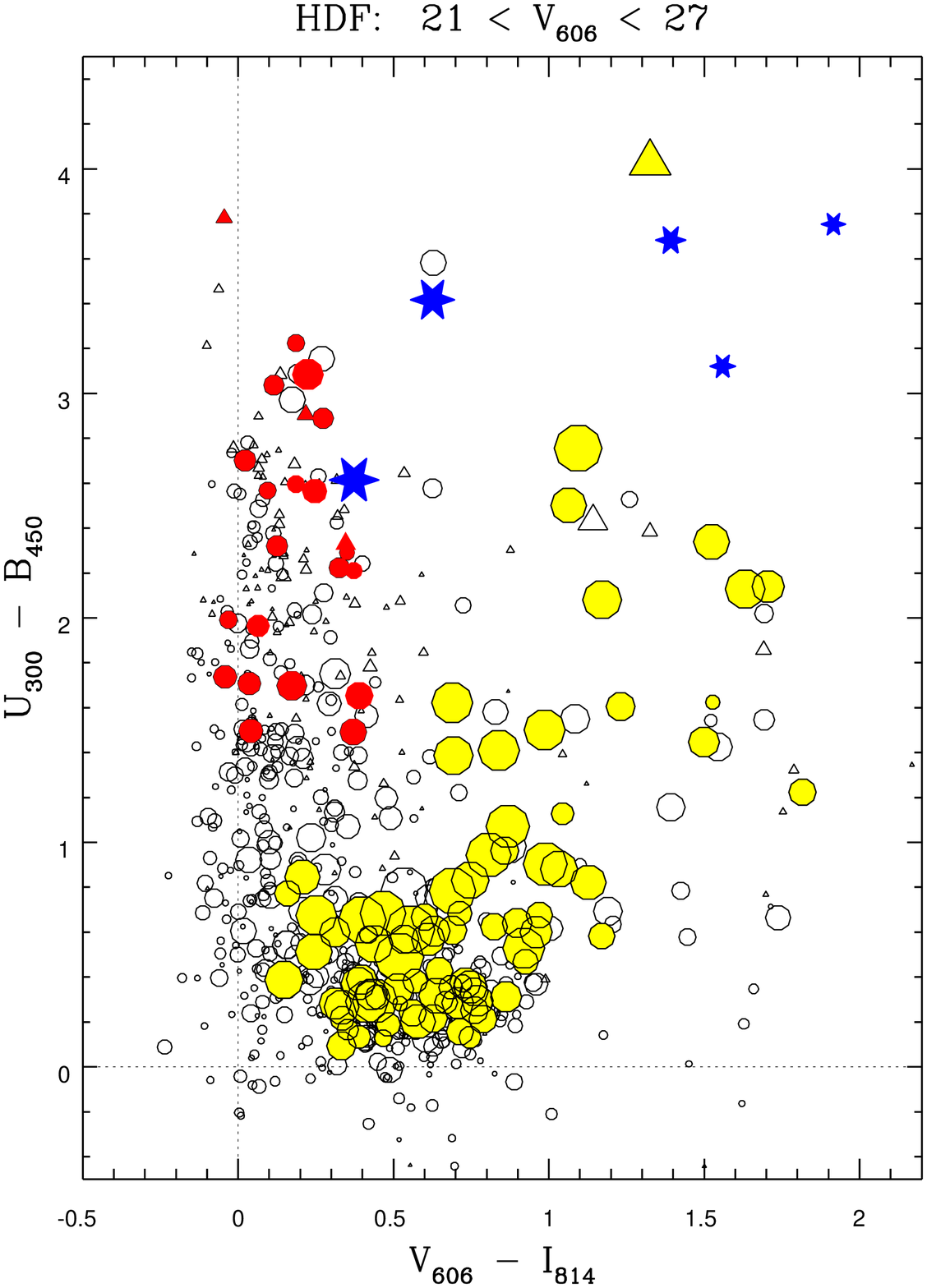,height=6in}}
\caption{Another multi--color view of the HDF.  Here, the horizontal
axis plots the $\V606 - \I814$ color, which is relatively insensitive 
to intergalactic Lyman~$\alpha$ absorption for redshifts $z < 4$.   
In this particular multi--color space the ``plume'' of high redshift 
galaxies stands nearly vertical, without the tilt seen in figure 2.   
Symbols are the same as in figure 2.}
\end{figure}

The F300W filter in WFPC2 is substantially bluer than $U$ bandpasses
used at terrestrial observatories.   As such, \U300 photometry is
sensitive to the passage of the Lyman break at significantly lower
redshifts than is the case for, e.g., the $U_n G {\cal R}$ system
used by Steidel et al.\  For example, roughly 90\% of galaxies in
our ground--based $U_n G {\cal R}$ survey lie in the redshift range
$2.6 < z < 3.4$.   By $z=2$, however, the Lyman limit is already 
well into the WFPC2 F300W bandpass and produces a recognizably
large color signature.  Followup spectroscopy has already identified 
\U300--dropout galaxies down to $z = 2.01$, and others which did not 
successfully yield redshifts quite likely lie at redshifts slightly 
below 2, where the absence of strong spectral features in the wavelength 
range presently accessible to LRIS at Keck makes secure redshift 
measurement difficult.  The commissioning of a new blue channel on 
LRIS will soon make it possible to measure redshifts for many more 
\U300 dropouts in the HDF.

The sensitivity of the \U300 filter to Lyman break galaxies at
lower redshifts has benefits and disadvantages.  It opens a larger
volume of redshift space to Lyman break selection, thus increasing
the number of objects which may be identified this way.  At the same 
time, as already noted, it makes many of them more difficult to confirm 
with optical spectra.  However, the ease of selecting galaxies at 
$2 < z < 2.5$ has one spectroscopic advantage: this is the range of 
``magic redshifts'' for near--infrared spectroscopy, where the 
[OII], [OIII], H$\beta$ and H$\alpha$ lines are all accessible
in the $J$, $H$ and $K$ bands.  This may make the HDF Lyman break 
galaxies particularly useful for studying nebular emission line 
diagnostics;  already Elston et al.\ have observed H$\alpha$ from
several HDF Lyman break galaxies at $z \approx 2$ (see below).

Various groups have created photometrically defined samples of
high redshift galaxy candidates in the HDF.   The color
selection criteria differ, and no ``definitive'' method has yet been
established.  The number of Lyman break candidates therefore varies
from sample to sample depending on the criteria which are adopted.
For example, Madau et al.\ (1996) defined conservative selection 
criteria based on models of galaxy color distributions in order
to select $z > 2$ galaxies while avoiding significant risk of 
contamination from objects at lower redshift.  These criteria,
however, miss some of the galaxies which are now spectroscopically 
confirmed to have $z > 2$.   Using the wealth of redshift information
now available in the HDF we can revisit this question and refine the
selection criteria.  As it happens, applying the same criteria used
by Steidel et al.\ for selecting Lyman break galaxies in the 
ground--based $U_n G {\cal R}$ color system does an excellent job of 
isolating $z > 2$ galaxies in the HDF.  Specifically, we may use 
the so--called ``marginal'' criteria of Steidel et al., namely
$$\U300 - \B450 \ge \B450 - \V606 + 1.0, \hspace{0.5cm} \B450 - \V606 \le 1.2.$$
This successfully recovers all 24 spectroscopically confirmed Lyman
break galaxies in the HDF at $2 < z < 3.5$.  As is the case for the 
ground--based sample, the only substantial contaminants are galactic stars,
some of which also satisfy these criteria, but these are easily
recognized in the WFPC2 images and excluded.  (All obvious stars
with these colors have also already been observed spectroscopically,
as it turns out.)  The only known low--redshift galaxy which satisfies 
these criteria is, of all things, the brightest galaxy in the HDF:  
an elliptical in WF2 at $z=0.089$.  This is not surprising, as 
ellipticals have colors similar to those of K--stars, and are thus 
``interlopers'' for the same reason until their redshifts become
large enough ($z \approx 0.1$) for $k$--correction effects to move
them out of the color selection region.    There should be very 
few $z < 0.1$ early--type galaxies in the small volume of the HDF which 
populate this part of color--color space.\footnote{For those particularly 
worried about foreground contamination, an additional color cut, 
$\V606 - \I814 \leq 0.5$, should successfully exclude most interlopers 
at $z < 0.1$.}  

Excluding stars and the few obvious $z < 0.1$ interlopers, there are 
approximately 187 galaxies in the HDF with $\V606 < 27.0$ which satisfy 
these color criteria and thus probably fall in the redshift range 
$2 \simlt z \simlt 3.5$.  The exact number depends somewhat on the choice 
of what constitutes a single galaxy with multiple clumps versus separate 
objects with small angular separations.  E.g. should the ``quad galaxy,'' 
4--858.0 at $z=3.22$, be considered a single galaxy with four pieces or four 
separate objects?  Such distinctions may be merely semantic or may be very 
important depending on the question being asked (cf. Colley et al.\ (1996)).

\subsection{HDF high--$z$ roundup}

To date, 26 galaxies in the HDF have spectroscopic redshifts $z > 2$.  
23 of these were pre--selected as Lyman break objects for spectroscopic
observation.   One galaxy, 4-445.0, was observed by Cohen et al.\ (1996) 
as part of a magnitude limited spectroscopic sample, but qualifies 
photometrically as a Lyman break galaxy nevertheless.  One other was 
was observed serendipitously (2-585.1 -- see below), but also satisfies 
the Lyman break color criteria defined above.  Finally, one galaxy 
(3-577.0) was observed as a gravitational lens candidate;  it is too 
faint for a robust \U300 Lyman break measurement and so does not qualify 
(see below).  Table 1 lists these objects, including several galaxies 
from our own observations which are previously unpublished.  There are 
many more Lyman break candidates, primarily $U_{300}$ dropouts, which 
have spectroscopically accessible magnitudes, and with considerable 
effort this list could eventually more than double in length.  Few of 
the $B_{450}$ dropout objects ($z \sim 4$ candidates) are bright enough 
to tempt the spectroscopist, but if they have strong emission lines 
a few may still yield redshifts.

\tabcolsep 0.1in 
\begin{table}
\center{HDF Galaxies with Spectroscopic Redshifts $z > 2$}
\begin{tabular}{lcclcl}
\multicolumn{1}{c}{ID} & RA & Dec & \multicolumn{1}{c}{$z$} & $V_{606}$ & Reference \\
\hline
2-449.0  &  12:36:48.332  &  62:14:16.67  &  2.008  &  23.68  &  S \\
2-585.1  &  12:36:49.808  &  62:14:15.18  &  2.008: &  23.85  &  E \\
3-118.1  &  12:36:54.724  &  62:13:14.74  &  2.232  &  24.41  &  * \\
2-903.0  &  12:36:55.071  &  62:13:47.05  &  2.233  &  24.60  &  L \\
2-525.0  &  12:36:50.120  &  62:14:01.04  &  2.237  &  24.80  &  * \\
2-82.1   &  12:36:44.077  &  62:14:09.98  &  2.267  &  24.54  &  L \\
4-445.0  &  12:36:44.637  &  62:12:27.34  &  2.268  &  24.08  &  C \\
2-824.0  &  12:36:54.617  &  62:13:41.28  &  2.419  &  25.23  &  L \\
2-239.0  &  12:36:45.883  &  62:14:12.10  &  2.427  &  24.54  &  * \\
2-591.2  &  12:36:53.175  &  62:13:22.76  &  2.489  &  24.91  &  * \\
4-639.1  &  12:36:41.715  &  62:12:38.84  &  2.591  &  24.74  &  S \\
4-555.1  &  12:36:45.344  &  62:11:52.67  &  2.803  &  23.41  &  S \\
1-54.0   &  12:36:44.094  &  62:13:10.75  &  2.929  &  24.44  &  * \\
4-52.0   &  12:36:47.687  &  62:12:55.98  &  2.931  &  24.37  &  L \\
4-289.0  &  12:36:46.944  &  62:12:26.09  &  2.969  &  25.17  &  * \\
4-363.0  &  12:36:48.297  &  62:11:45.88  &  2.980  &  25.05  &  L \\
2-643.0  &  12:36:53.427  &  62:13:29.38  &  2.991  &  24.87  &  L \\
2-76.11  &  12:36:45.357  &  62:13:46.98  &  3.160  &  25.32  &  L \\
2-565.0  &  12:36:51.186  &  62:13:48.79  &  3.162  &  25.17  &  * \\
2-901.0  &  12:36:53.607  &  62:14:10.25  &  3.181  &  24.87  &  L \\
4-858.0  &  12:36:41.233  &  62:12:03.00  &  3.216  &  24.28  &  S,L,Z \\
3-243.0  &  12:36:49.817  &  62:12:48.88  &  3.233  &  25.60  &  L \\
3-577.0  &  12:36:52.247  &  62:12:27.18  &  3.360  &  27.25  &  Z \\
2-637.0  &  12:36:52.747  &  62:13:39.08  &  3.368  &  25.27  &  L \\
2-604.0  &  12:36:52.407  &  62:13:37.68  &  3.430  &  25.25  &  L \\
3-512.0  &  12:36:56.117  &  62:12:44.69  &  4.022  &  25.85  &  * \\
\hline
\multicolumn{6}{c}{\bf References} \\
\multicolumn{6}{c}{S: Steidel et al.\ 1996b;  L: Lowenthal et al.\ 1997; C: Cohen et al.\ 1996;} \\
\multicolumn{6}{c}{Z: Zepf et al.\ 1997; E: Elston et al.\ private communication;} \\
\multicolumn{6}{c}{*: this paper (Steidel et al.\ observations, unpublished)} \\
\\
\end{tabular}
\end{table}

Here are a few notes on individual Lyman break galaxies, including
corrigenda to some errors in our first paper on HDF spectroscopy
(Steidel et al.\ 1996b).

\vspace{0.7cm}

\begin{description}

\item[2-449.0:~~~~] This galaxy was erroneously identified as having 
$z = 2.845$ in Steidel et al.\ 1996b (object C2-05 in that paper).  The broad 
band spectral energy distribution of the galaxy, both in the UV and near--IR,
appears to be inconsistent with that redshift (e.g. the Lyman break amplitude
is too small for the galaxy to be at $z=2.8$).  We reanalyzed the spectrum 
and found that $z=2.008$ is more likely to be the correct redshift.  The 
galaxy was then observed by Elston et al.\ in the infrared to search for 
H$\alpha$ emission, both with narrow band imaging at the IRTF and with 
the Cryogenic Spectrograph (CRSP) at KPNO.  Both observations detected 
strong line emission, confirming the new redshift.

\item[2-585.1:~~~~] This galaxy, a $U_{300}$ Lyman break galaxy with 
a particularly dramatic morphology (see figure~15 below), lies several 
arcseconds away from 2-449.1 (see above).  In their CRSP observations of 
2-449.1, Elston et al.\ also detected line emission from 2-585.1 which 
fortuitously fell on their spectrograph slit.  The line was confirmed 
with a subsequent observation targeting 2-585.1 itself.  Presuming
the line to be H$\alpha$, the redshift is approximately the same as that 
of 2-449.1.  Both galaxies have very similar $U_{300} - B_{450}$ colors, 
again suggesting similar redshifts.

\item[3-550.0:~~~~] This galaxy, object C3-02 from Steidel et al.\ 1996b, is not 
included in the table above.  As with 2-449.0 (above), the colors are in most
respects not consistent with the published redshift ($z=2.775$);  the relatively
small \U300 - \B450 color and near--IR SED both suggest a lower redshift.
The original spectrum has very poor signal--to--noise, and we now feel that 
the published redshift is probably incorrect.  Reobservations of this galaxy
have thus far been inconclusive.  We suspect that the galaxy has $z \simlt 2$, 
making spectroscopic redshift confirmation difficult with LRIS at the 
present time.

\item[3-577.0:~~~~] This is the faintest object (with $V_{606} > 27$!) in 
the HDF with a reported redshift.  Hogg et al.\ (1996) proposed that this is 
part of a gravitational lens system consisting of a red foreground elliptical,
a large blue arc, and 3-577.0 as a suggested counterimage of the arc.  
Zepf et al.\ (1997) observed the system and detected a weak emission line 
from 3-577.0 which they interpret as Lyman~$\alpha$.  Their spectra did not 
yield redshifts for the other components of the system.  Although the emission
line is very weak and the proposed redshift should perhaps be regarded as 
tentative, it is plausibly correct.  3-577.0 is too faint to be robustly 
measured as a ``dropout,'' as the $U_{300}$ photometry is not deep 
enough to set a strong enough constraint on the $U_{300} - B_{450}$ color.
However, the colors are not {\it inconsistent} with it being at $z \sim 3$.  
Moreover, photometrically the ``arc'' (3-593.0) does qualify as a Lyman 
break galaxy, and probably has $z > 2$.  Thus it remains plausible that
the two objects are one and the same, gravitationally lensed.
Arguing against this is the absence of any emission line in Zepf et al.'s
spectrum of the arc, and the fact that the two objects have apparently
different $B_{450} - V_{606}$ colors (although they are virtually identical 
at $V_{606} - I_{814}$).  To explain this would require that the long
slit observation, which crossed the arc perpendicularly, must simply have
missed the line emitting region of the magnified galaxy.  The $B_{450}-V_{606}$ 
colors, if the galaxies are at $z = 3.36$, could conceivably be affected by
different amounts of foreground Lyman~$\alpha$ forest absorption, while
the $V_{606}-I_{814}$ colors are free of this effect.  Further observations 
of this system are warranted.

\begin{figure}
\centerline{\psfig{figure=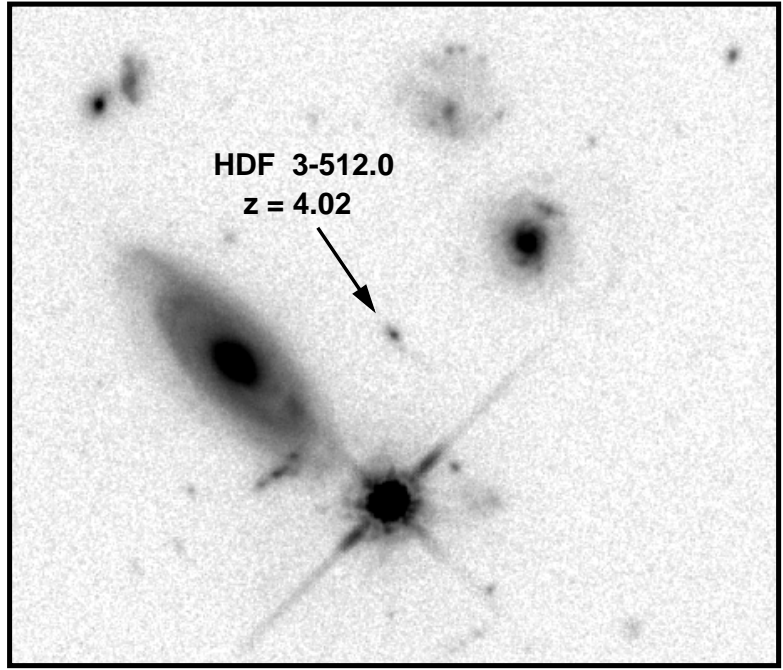,height=3.5in}}
\caption{Image of 3-512.0, one of the brightest \B450--dropout galaxies
in the HDF.   A single emission line detected in its spectrum suggests
that 3-512.0 has $z = 4.02$. }
\end{figure}

\begin{figure}
\centerline{\psfig {figure=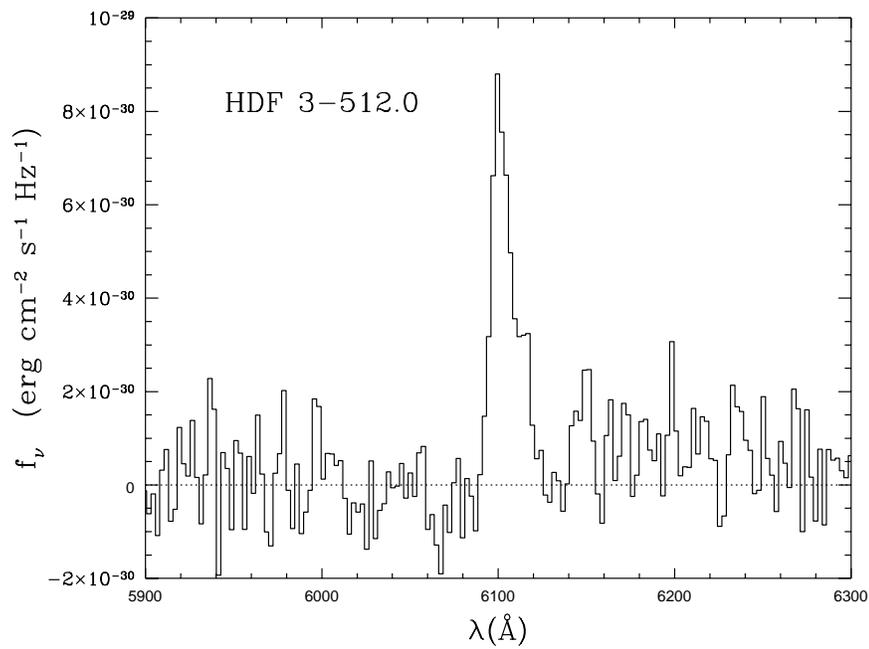,height=3.5in,angle=-90}}
\caption{Spectrum of 3-512.0, showing the emission line detected
at $\lambda 6105$\AA.  The asymmetric line profile and the continuum
discontinuity across the line are both consistent with this being
Lyman~$\alpha$ at $z = 4.02$.}
\end{figure}

\item[3-512.0:~~~~] Of the candidate $z \approx 4$ ``$B_{450}$--dropouts''
tabulated in Madau et al.\ 1996, only three objects have $I_{814} < 26$.
3-512.0 is the brightest of these at $V_{606}$.  The galaxy is in fact 
detected in the $B_{450}$ image, and does not drop out completely.   
If this is indeed a high redshift object, then the Lyman limit has only 
partially passed through the filter bandpass, suggesting $z \approx 4$.  
We observed this galaxy with LRIS at the Keck Observatory in May 1996 and 
again in March 1997.  Both observations detected a single emission line 
at $\lambda 6105$\AA.  If the emission line is identified with Lyman~$\alpha$,
this would imply a redshift $z=4.02$.  No other significant lines were 
detected.  In particular, if the observed line were [OII] $\lambda 3727$\AA\ 
or [OIII] $\lambda 5007$\AA\ from a low--redshift object our spectrum 
would cover the wavelength ranges where other strong optical lines are
expected, and none are seen.  The coincidence between 
color--selection as a ``weak'' $B$--dropout and the redshift derived
on the assumption that the line is Lyman~$\alpha$ is suggestive.
The spectrum (see figure 6) shows a flux discontinuity across 
the emission line, as would be expected for a $z \approx 4$ 
galaxy due to intervening absorption from the Lyman~$\alpha$ forest.  
Moreover, the emission  line profile is asymmetric, with a sharp
blue side and extended red wing.   This characteristic profile is often 
seen in high redshift quasars and galaxies (cf. the $z = 4.9$ galaxy of 
Franx et al.\ (1997)), where it is due to absorption of the blue side 
of Lyman~$\alpha$ by intervening neutral hydrogen.   Although 
a single--line redshift cannot be regarded as 100\% secure, the additional 
circumstantial evidence suggests that this is indeed a galaxy at $z = 4.02$.

\end{description}

\vspace{0.5cm}

\begin{figure}
\centerline{\psfig {figure=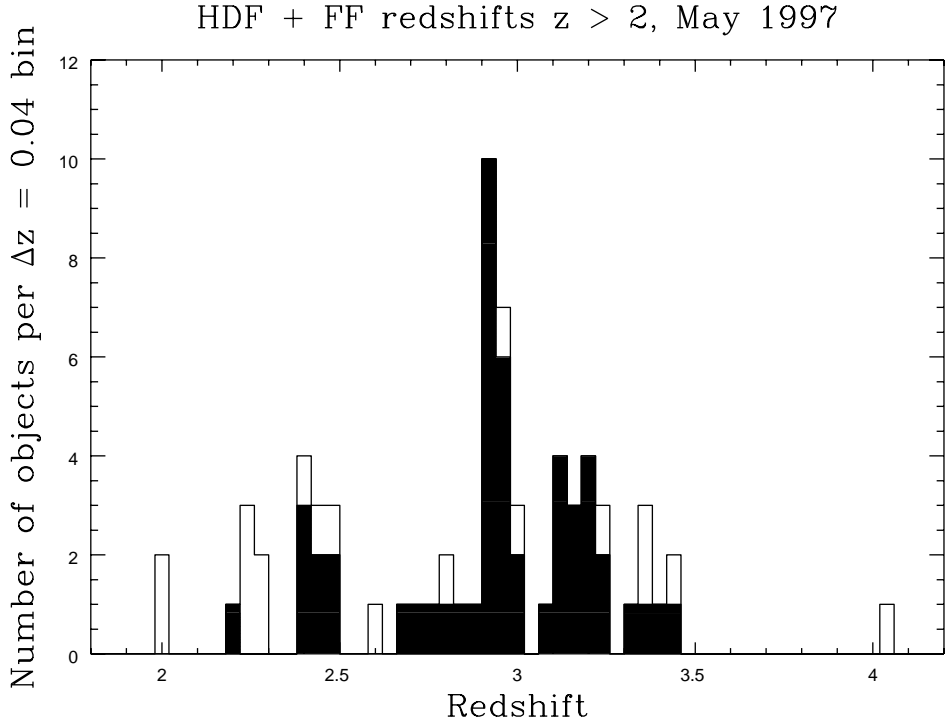,height=4in,angle=-90}}
\caption{The overall redshift distribution of galaxies with $z > 2$ in 
the HDF and its Flanking Fields.  The filled portion of the histogram 
are galaxies selected from our ground--based $U_n G {\cal R}$ imaging, 
primarily outside the central HDF, while the open histogram includes
additional Lyman break galaxies in the central HDF.}
\end{figure}

As part of our general survey for Lyman break galaxies 
we have used the Palomar 200--inch telescope to image a wider region 
(roughly $9\arcmin \times 9\arcmin$) which includes the HDF and its flanking 
fields, selecting candidates using our standard $U_n G {\cal R}$ color 
criteria.  The overall redshift distribution of $z > 2$ galaxies now known 
from the central HDF and the surrounding region is shown in figure 7.   
As has been often noted from surveys at $z < 1$, the redshift distribution 
in this narrow pencil beam is highly non--random, with prominent ``spikes''
where the galaxy density is large.  The existence and significance
of strong clustering in the Lyman break galaxy population is discussed 
in several papers now in press:  Steidel et al.\ (1998) report on a strong 
redshift ``spike'' seen in another survey field, while Giavalisco et al.\ 
(1998;  see also this volume) have measured the angular correlation
function of Lyman break galaxies.   Figure 7, as well as data we have
collected in other fields, shows that these highly overdense 
redshift--space structures are ubiquitous at $z \approx 3$ -- the
case studied in Steidel et al.\ (1998) is not unique, and we find 
similar spikes in essentially all of our survey fields
once sufficiently large numbers of redshifts have been collected.
The HDF is no exception.

\section {Statistics of Lyman break galaxies}

\begin{figure}
\centerline{\psfig {figure=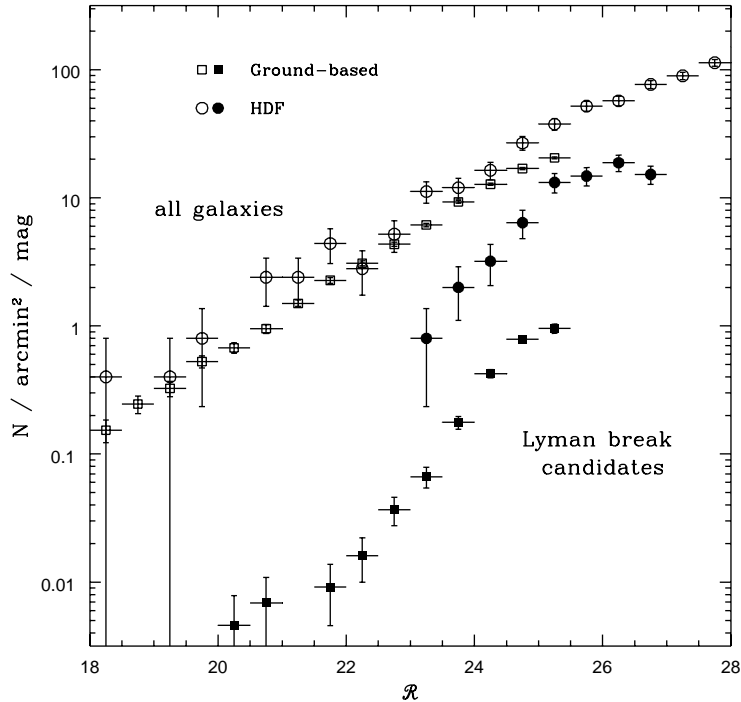,height=4in}}
\caption{Number counts of Lyman break galaxies from ground--based
and HDF samples compared to the overall counts of faint field galaxies.
$R$ magnitudes are approximated for HDF galaxies as $(\V606 + \I814)/2.$}
\end{figure}

What fraction of the faint galaxy population lies at these large
redshifts?  This was the question addressed by Guhathakurta et al.\ (1990),
who were among the first to apply the Lyman break technique to faint 
galaxy samples.  At the magnitude limits of their sample they found
few $U$--band dropout objects and concluded that only a small fraction 
of faint galaxies lie at $z \approx 3$.    However this fraction
rises as we go fainter;  the exact proportion depends on the color
selection (different criteria probe different volumes at high
redshift) and the photometric depth of the sample.  In figure 8
I show the overall number counts of galaxies in our ground--based
survey and in the HDF, compared to the number of Lyman break candidates 
in each.  Brighter than ${\cal R} = 22.9$, all of the color--selected 
objects in the ground--based sample have turned out to be galactic 
stars (plus a small number of high redshift QSOs) -- we have yet to 
find a Lyman break galaxy with ${\cal R} < 22.9$ in the 0.24 square 
degrees which we have surveyed.  Fainter than this, the number of 
Lyman break objects rises rapidly, reaching 5\% of the galaxy population 
at ${\cal R} = 25.0$.  In the HDF the counts of Lyman break galaxies
are higher.   This is partially due to the larger volume probed by 
the color selection technique in the HDF because of the bluer F300W 
filter.   It may also reflect genuine redshift evolution in the galaxy 
population, however, since the HDF \U300--dropout galaxies have lower 
mean redshifts than do the objects in the ground--based sample, and
the total UV luminosity density in galaxies is evidently rising
from $z \approx 4$ to 2 (Madau et al.\ 1996).  At $R \approx 26.5$, 
nearly 1 in 4 galaxies in the HDF is probably at $z \simgt 2$.

With a robust technique for identifying large numbers of galaxies
in a particular redshift range we may quantify various statistical 
properties of the population even without complete spectroscopic 
redshift information.  Here I will consider luminosity and 
color distributions of Lyman break galaxies.  Another example
is the angular correlation function, which is addressed in the 
contribution of Giavalisco to this volume.  

\subsection{Luminosity functions}

The redshift selection function of our ground--based $U_n G {\cal R}$
survey is now well categorized, with more than 400 spectroscopic 
redshifts measured in many independent survey fields.  
For a particular color--defined subset of our sample
we find that approximately 90\% of the galaxies lie at $2.6 < z < 3.4$.  
The front--to--back ``depth'' of this redshift range is small, 
photometrically speaking (i.e. in terms of distance modulus), and thus 
the counts of Lyman break galaxies, even without confirming redshifts, 
primarily reflect their intrinsic luminosity distribution.  

The physical significance of the ``luminosity function'' of Lyman break 
galaxies is different from that of the more familiar optical luminosity 
functions that have been determined locally and out to $z \approx 1$ 
from the CFRS, AUTOFIB, CNOC and Hawaii Surveys.   By observing 
$z \approx 3$ galaxies through optical bandpasses we are measuring their 
luminosities at rest--frame ultraviolet wavelengths of approximately
1500\AA.  For young galaxies, ultraviolet continuum emission arises 
mainly from hot, massive stars, modulated by the absorbing effects 
of dust.  In the absence of extinction, the ultraviolet luminosity 
thus primarily reflects an {\it instantaneous} property of a galaxy:
its star formation rate.  The UV luminosity declines rapidly after 
the cessation of star formation as the O and B stars which produce 
it burn off the main sequence.  In more local galaxy samples, the 
rest--frame optical light used to define luminosity functions manifests
some integral over the past star formation history of a galaxy, and thus 
better describes its total stellar content.  It is therefore most 
straightforward to interpret the UV luminosity function (UVLF) of 
Lyman break galaxies as a distribution of star formation rates in 
the population.  The complication, however, is that the effects
of extinction on UV emission can potentially be large, and are
at present mostly unknown for $z \approx 3$ galaxies.  I return to
this issue below, considering only the ``raw'' luminosities here.

Figure 9 presents a composite luminosity distribution for Lyman 
break galaxies derived from the ground--based and HDF samples.  
Here I briefly describe the procedures used in constructing this 
diagram in order to point out some of the inherent uncertainties.  
A detailed discussion and analysis will be presented in a forthcoming 
paper (Dickinson et al.\ 1998).

The overall luminosity distribution has been normalized using data from 
our ground--based sample.  Here the statistics are good thanks to the
large number of Lyman break objects and the use of many survey fields to
average over local fluctuations, and the redshift selection function 
is well characterized through extensive spectroscopy.  
At brighter magnitudes (${\cal R} < 24.5$) 
there is significant contamination from galactic stars (plus a few QSOs).  
At those magnitudes, therefore, only spectroscopically confirmed galaxies 
have been used.  From $24.5 < {\cal R} < 25.5$, the number counts from 
Figure 9 are used with a small, modeled correction for incompleteness.
Spectroscopy has confirmed that the stellar and QSO contamination 
is very small at these magnitudes.  The measured 
redshift selection function of the ground--based sample is used to 
normalize the survey volume.  The true normalization is probably 
higher, as I have assumed that the color criteria are 100\% efficient 
at the peak of the selection function ($z = 3.0$).  Comparison of
our ground--based Lyman break sample in the HDF and the WFPC2 images
demonstrates that we do miss some fraction of candidates, even at
relatively bright magnitudes, due to photometric confusion with
foreground galaxies.  For example, 4--555.1, a galaxy at $z = 2.803$
which is one of the brightest Lyman break objects in the HDF (and which
is illustrated in figure~1 above) is missed in the ground--based sample 
due to flux contamination by a foreground elliptical galaxy only 
$\sim 2\arcsec$ away.

\begin{figure}
\centerline{\psfig {figure=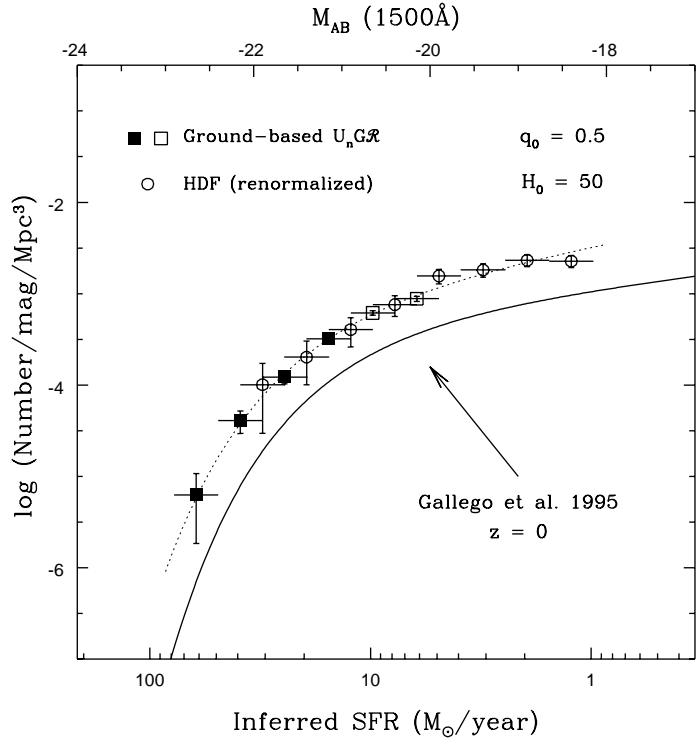,height=4in}}
\caption{The ultraviolet luminosity function of galaxies at $z \approx 3$.  
Filled and open squares are derived from spectroscopically and 
photometrically selected galaxies from the ground--based survey, 
while circles are renormalized counts of objects from the HDF.
Absolute magnitudes are computed on the AB system: 
$M_{AB} = -21$ corresponds to a specific luminosity of 
$1.1 \times 10^{29}$~erg~s$^{-1}$~Hz$^{-1}$.  The local H$\alpha$
luminosity function of Gallego et al.\ (1995) is shown for comparison.
Both UV and H$\alpha$ luminosities are converted to star
formation rates (bottom axis) using a Salpeter IMF.  See text for 
discussion.}
\end{figure}

The ground--based data, reaching ${\cal R} = 25.5$, probes only the
relatively bright end of the Lyman break galaxy luminosity function.
The HDF offers an opportunity to sample lower luminosities.
The difficulty with the HDF sample, however, is that its redshift
selection function is quite uncertain;  only 24 $\U300$--dropouts
have measured redshifts.   Moreover, the spectroscopic success rate 
is probably declining rapidly at $z < 2.3$ due to the lack of strong 
spectral features in the wavelength range currently accessible to LRIS.   
Finally, there is only one HDF and it covers a {\it very} small region 
of the sky.  As noted above, we now know that Lyman break galaxies 
are strongly clustered, and thus any single, narrow pencil beam survey 
will encounter a highly non--random galaxy distribution.
This complicates the determination of the redshift selection function, 
and may compromise any attempt to normalize the luminosity function 
simply because field--to--field variations cannot be averaged away.
The Southern HDF, planned for 1998, will offer a second ``realization'' 
of the F300W dropout sample for comparison, but still the total volume 
surveyed will be small.  

In addition, as noted above, the redshift range of the HDF Lyman 
break galaxies is much broader than that of the ground--based sample,
and extends to lower redshifts.  HDF \U300--dropouts with measured 
redshifts have $\langle z \rangle \approx 2.7$, but it is 
likely that the mean redshift of the complete photometrically selected 
sample is lower still.     Madau et al.\ (1996) suggest that the UV 
luminosity density of the universe, an integral over the luminosity 
function, was rising steeply with time at this cosmic epoch, a result 
derived from the HDF data by comparing numbers of \U300-- and \B450--dropout 
objects.  Therefore the luminosity function itself may have evolved
rapidly at these redshifts, providing an additional uncertainty
for splicing the ground--based and HDF Lyman break samples together.
% The difference in the mean redshifts of the sample is small, however,
% representing $\simlt 2$\% of the age of the universe, or an interval 
% of $< 0.3$ Gyr, so this may not be a dominant uncertainty.
Also, this broader redshift range means that the photometric ``depth'' 
of the survey (in terms of relative distance modulus, front to back) 
is larger.  This complicates the transformation from apparent magnitude
distribution to luminosity function, both by blurring the distance modulus 
conversion and by making k--correction effects somewhat larger.

For the present purposes, I have used the HDF sample mainly to provide 
an indication of the UVLF slope at faint luminosities.  Magnitudes of HDF 
galaxies are transformed to luminosities assuming $\langle z \rangle = 2.6$.  
Initially we normalize the survey volume by assuming unit selection
efficiency over the range $2 < z < 3.5$.  The resulting HDF luminosity
function is still higher than the ground--based counts over the range
of luminosity overlap.   In part this may be due to incompleteness 
in the ground--based sample, but it is likely that it also manifests
to the effect of redshift evolution.  Here, the HDF space densities
have been scaled downward to match that of the ground--based sample 
in the luminosity range of overlap.   The faintest data points 
in figure 9 should be regarded with caution, as the sample may suffer 
as--yet unquantified incompleteness effects at its photometric limits.  
Thus while the luminosity function appears to be flattening, 
the apparent slope should be taken as a limit to the true value.

One measure of the distribution of galaxy star formation rates 
in the local universe is the H$\alpha$ luminosity function of 
Gallego et al.\ (1995).    In figure 9, both the $z \approx 3$ UVLF
and the H$\alpha$ measurements at $z \approx 0$ have been converted 
to star formation rates (SFR) using consistent assumptions of a 
Salpeter IMF spanning 0.1--125 $M_\odot$ (e.g. Madau et al.\ 1998;
adopting the Kennicutt 1983 conversion for H$\alpha$, which assumes
a somewhat different IMF, increases the Gallego et al.\ SFR values by 26\%).  
The SFR distribution at $z = 3$ is strikingly like that measured
at $z = 0$, spanning a similar range but with more galaxies per unit 
volume forming stars at any given rate.  The characteristic ``$L^\ast$'' 
of the $z = 3$ and $z = 0$ SFR distributions are approximately 
14 and 10 $M_\odot$~yr$^{-1}$, respectively, and the faint end slopes 
are similar.    

These similarities may very well be coincidental.  No correction has 
been made for the effects of extinction in the $z = 3$ galaxies 
(the Gallego et al.\ H$\alpha$ data does include an extinction correction).  
As we will see below, these corrections could be quite significant.   
Also, the Lyman break luminosity function in figure 9 is plotted for an 
Einstein--de Sitter cosmology.  For an open universe the $z \approx 3$ 
data translates downward and to the left relative to the local data, 
i.e. toward higher luminosities/SFRs but lower space densities.  
Thus both extinction and cosmology could work in the direction
of increasing the typical star formation rates inferred for high
redshift galaxies.

The physical state of star formation in the distant galaxies may be 
quite different than that in the ``typical'' galaxy actively forming stars
in the nearby universe.  The sizes of Lyman break galaxies are much smaller 
than those of ordinary galaxies with similar star formation rates nearby 
(Giavalisco et al.\ 1996b, Lowenthal et al.\ 1997), and their ultraviolet 
surface brightnesses are much higher, comparable to those of powerful 
starburst galaxies today (Meurer et al.\ 1997,  Giavalisco et al.\ 1996a).   
Again, this suggests that the similarity of the SFR distributions at 
$z \approx 0$ and 3 may be, to a certain extent, a coincidence.

The integral over the best fit to the UVLF shown in figure~9 
gives a comoving 1500\AA\ luminosity density at $z \approx 3$ of 
$2.1 \times 10^{26}$~erg~s$^{-1}$~Hz$^{-1}$~Mpc$^{-3}$, corresponding to 
a star formation density (using the same Salpeter conversion) of 
$0.026 M_\odot$~yr$^{-1}$~Mpc$^{-3}$, again neglecting any correction
for extinction.  This value should be regarded as a lower limit because 
of the possibility of incompleteness in the ground--based sample,
although comparison between WFPC2 and ground--based Lyman limit samples
suggests such incompleteness is not likely to exceed 30\%.  Varying 
the procedure used to construct the data set changes the form of the 
luminosity function somewhat, and the integrated luminosity
density varies from 1.6 to $3.5 \times 10^{26}$~erg~s$^{-1}$~Hz$^{-1}$~Mpc$^{-3}$
because of these systematic changes -- this is the dominant uncertainty,
significantly exceeding shot noise in the data.  As noted above, 
the integrated luminosity density of the HDF Lyman break galaxies appears to 
be somewhat larger than that implied from the $z \approx 3$ UVLF which
is normalized by the ground--based sample, but uncertainties about the 
redshift selection function of the HDF sample make it hard to estimate 
by how much.  If real, the excess could be the consequence either of 
clustering in the small HDF volume or of redshift evolution of the galaxy 
population, with a larger luminosity density present at the lower redshifts 
probed by the HDF Lyman break color selection.

\subsection{Colors of high redshift galaxies and measures of extinction}

The preceding discussion of the UVLF of Lyman break galaxies and 
their star formation explicitly neglected the effect of extinction,
which could be strong at ultraviolet wavelengths even if the dust
content of these galaxies is relatively small.  Indeed we are
reasonably sure that these objects do contain some amount
of dust, as their Lyman~$\alpha$ emission lines are generally
much weaker than expected from their UV--derived star formation
rates under the assumption of Case B recombination (Steidel et al.\
1996a).  Lyman~$\alpha$, however, is a resonance line and is
easily extinguished with small amounts of dust, so there is little
constraint on the dust content from this spectral feature.

A dust--free, star forming galaxy should have a very blue
UV continuum -- flat in $f_{\nu}$ units if star formation has 
been proceeding for $\simgt 10^8$ years, or even bluer for very 
young starburst populations.  If we examine the actual UV
spectral slopes of Lyman break galaxies, however, we find that
they are mostly redder than flat spectrum in $f_{\nu}$.
Figure~10 plots the $G - {\cal R}$ color of more than 400 galaxies 
from our ground--based sample versus redshift.  For this color
combination knowing the redshift of each galaxy is important 
because Lyman~$\alpha$ forest opacity can effect the flux measured
in the $G$--band, making colors redder at larger redshifts
independent of extinction internal to the galaxy.  In figure 10,
the predicted colors of star--forming galaxies are plotted versus
redshift for various amounts of internal extinction.  
The unreddened models define the blue envelope of the color
distribution, with nearly all galaxies being redder than the
colors expected from a naked star forming galaxy.
This effect is also seen in local starburst galaxy samples
(e.g. Calzetti et al.\ 1994), and has been studied in Lyman
break galaxies by Meurer et al.\ (1997).  Although for some
objects this reddening of the UV spectral slope may be due
to ageing of a starburst with a declining star formation rate,
it is likely that some or most of the effect is indeed due
to extinction.  

\begin{figure}
\centerline{\psfig {figure=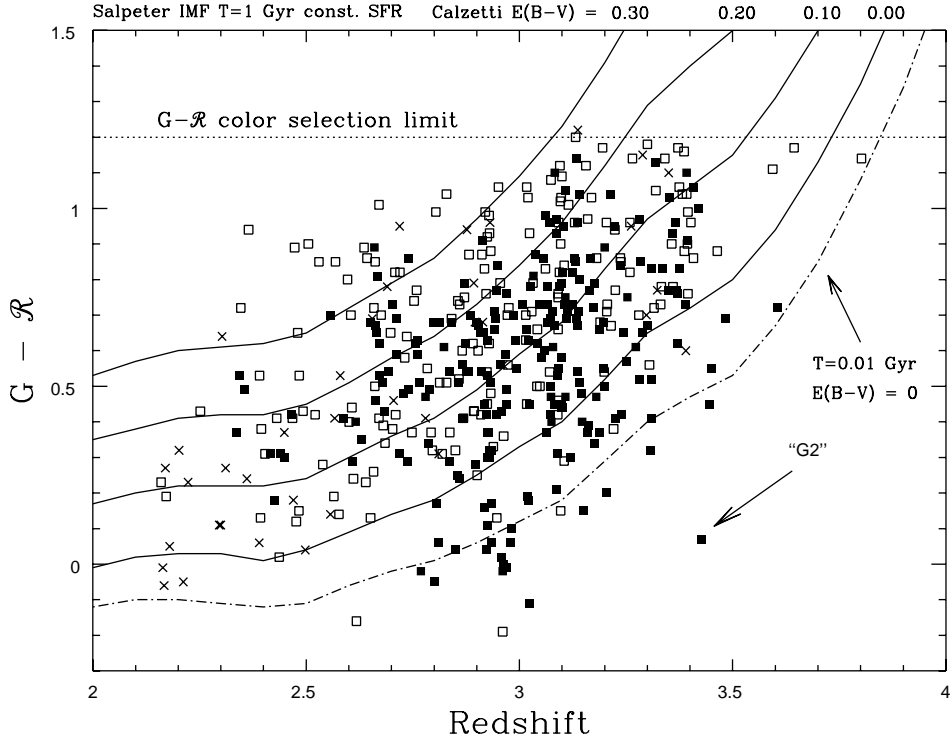,height=4in,angle=-90}}
\caption{$G - R$ color of Lyman break galaxies versus redshift.  The
various symbol types mark different color--selected subsamples and
are not important here.  The solid lines show the predicted colors
of actively star forming galaxies with various amounts of reddening,
using the Calzetti et al.\ (1994) starburst extinction law.  The bluest
model track shows an unreddened starburst model with very young 
($10^7$~year) age, i.e. O--star dominated and intrinsically bluer 
than flat spectrum.  The unreddened models define the blue envelope
of the color distribution.  Some objects which are bluer than the 
unreddened models have colors affected by strong Lyman~$\alpha$
emission lines in the $G$--band, e.g. the galaxy labeled G2 which
is an AGN with very strong line emission.}
\end{figure}

\begin{figure}
\centerline{\psfig {figure=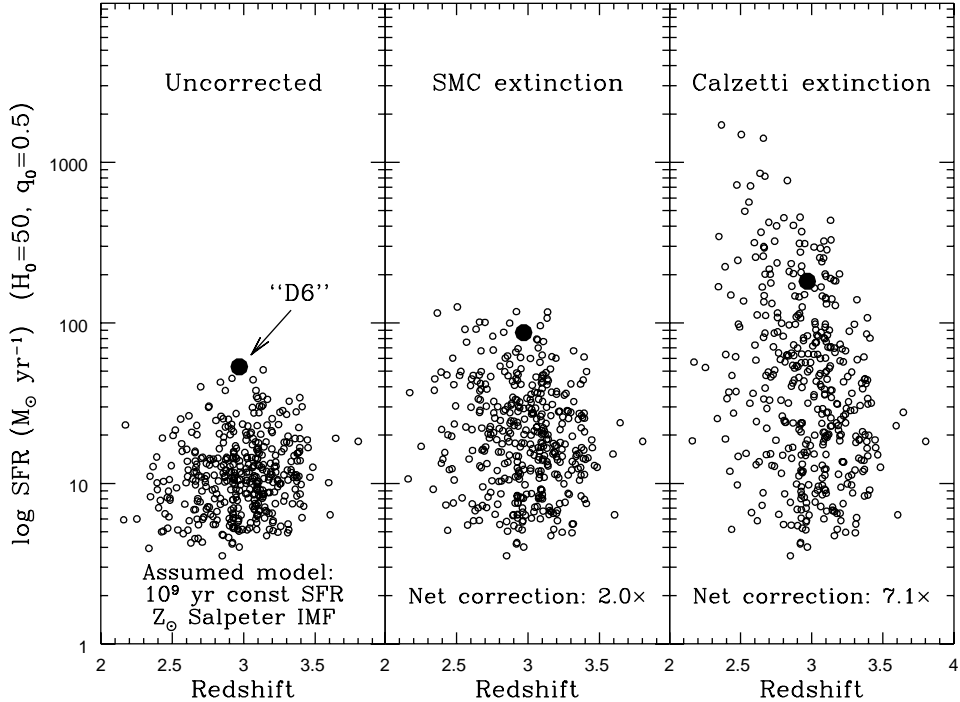,height=4in,angle=-90}}
\caption{Estimates of the effect of extinction on the UV luminosities
and derived star formation rates of Lyman break galaxies.  The left
panel shows computed star formation rates for objects with redshifts
in our survey.  The most luminous galaxy in the sample, labelled
D6, is shown with a large symbol for identification in each panel.
In the center and right panels, the $G - R$ color of each object has
been used to compute reddening using SMC and Calzetti extinction
laws and the appropriate correction has been applied to the star 
formation rates.  At bottom, the net correction to the {\it observed}
population of objects is tabulated.}
\end{figure}

The difficulty in interpreting UV spectral slopes as a measure
of extinction is that the inferred luminosity corrections are 
tremendously sensitive to the form of the reddening law at UV 
wavelengths.  For local starburst galaxies, an effective 
attenuation law has been derived which relates UV slope to
total extinction (cf. Calzetti et al.\ 1994;  Calzetti 1997).
The wavelength dependence of this attenuation law in the
near--UV is very ``grey,'' such that a small change in the
UV color requires a large change in the total extinction.
Varying the UV slope of the extinction law, as well as its
normalization, can dramatically change the derived total
suppression of UV galaxy luminosities.  This is illustrated
in figure 11, where I have used the observed colors of
Lyman break galaxies in our spectroscopic sample to infer
the UV extinction at 1500\AA\ under the assumption of two
dust attenuation laws:  the SMC extinction curve and the
local starburst extinction prescription of Calzetti (1997).
Without correction, the star formation rates derived for the
brightest galaxies in our sample are $\sim 50 M_{\odot}$/year.
With SMC extinction, they slightly exceed $100 M_{\odot}$/year,
and the net correction to the overall star formation rate of
the Lyman break population is a factor of two.  For the
starburst attenuation law, the most luminous galaxies approach
SFRs of $2000 M_{\odot}$/year, and the net correction to
the {\it observed} population of galaxies is a factor of 7.1;
the actual effect of dust on the global star formation rate
would be larger because some intrinsically luminous but reddened 
galaxies would disappear from of a flux--limited sample.
Meurer et al.\ (1997), using an extinction relation calibrated
for local starbursts using correlations between UV spectral slope
and far--infrared emission, derive even larger correction
factors of $\sim 15\times$ for the Lyman break population, 
in part due to different assumptions about the spectral slope 
of the underlying, unreddened continuum.

At present, we have little direct information about the true
effects of extinction on the UV luminosities of Lyman break galaxies.
While it is plausible that the attenuation law for local starburst
galaxies may apply to their high redshift ancestors, the inferred 
effect on the global UV luminosity is so sensitive to small 
variations in the extinction law that one wishes for independent 
data at other wavelengths which could be used to verify star formation 
rates.  In the future, far--infrared measurements of thermally 
reradiated dust emission in Lyman break galaxies may be possible 
from SIRTF or WIRE, or with 
sub--millimeter observations with the instruments like SCUBA.
In the meanwhile, we have begun a program of near--infrared
spectroscopy at UKIRT to measure Balmer line emission from Lyman break 
galaxies, and thus provide an independent measure of their star formation
rates which should be less sensitive to extinction.  This work
is painfully slow compared to optical multislit spectroscopy
for measuring redshifts, requiring night--long exposures on
one galaxy at a time.  Preliminary results are reported by Pettini 
et al.\ 1997, and suggest that the star formation rates derived from
the Balmer emission may be a few times larger than those inferred
from the UV continuum.  A much larger sample is needed before
we can make secure statements, but at least the problem is
addressable by observation.   Moreover, the same is now true at
low redshift.  Recently, Tresse \& Maddox (1997) have derived an 
H$\alpha$ LF for CFRS galaxies at $z \approx 0.2$, while Treyer 
et al.\ (1997) presented a near--UV LF for galaxies at similar 
redshifts.  Comparison of these two LFs, converted to star 
formation rates, suggests that the UV luminosity density is
underestimated of this local sample is suppressed by 
$\sim 1$ magnitude, presumably due to the effects of dust.

\section{Infrared properties}

One difficulty with the HDF WFPC2 data for studying galaxies at
$z > 1$ is that their optical rest--frame light is redshifted 
beyond the WFPC2 filter bandpasses.  The WFPC2 images thus provide
only an ultraviolet view of the $z > 1$ universe.   While this
has advantages for detecting and studying active star formation in 
very distant galaxies, it makes it difficult to compare their 
properties to those of objects in the nearby universe
at familiar rest--frame wavelengths.  We may address this problem
by collecting data on the nearby universe in the ultraviolet,
or by studying HDF galaxies in the near--infrared.

The HDF has been imaged in the infrared by several groups.  Len Cowie and 
colleagues used the CFH and UH 2.2m telescopes to obtain images of the central 
deep field in $J$ and an ``$H$+$K$'' notched filter, and have imaged a much 
wider surrounding region ($9^{\prime} \times 9^{\prime}$) to shallower depths.
Hogg et al.\ (1997) used the NIRC camera on Keck to obtain deep images
of two small ($\approx 40\arcsec \times 40\arcsec$) fields within the
HDF, and the Caltech group has also covered a wider surrounding region
with the Palomar 60--inch.  My collaborators and I imaged the central 
HDF using IRIM on the KPNO Mayall 4m telescope over the course of ten 
nights.  The field of view of IRIM is well matched to that of the WFPC2, 
providing easy coverage of the complete HDF.  Data was collected in the 
$J$, $H$ and $K$ bands, and reach formal $5\sigma$ limiting magnitudes 
in a 2\arcsec\ diameter aperture of $J=23.5$, $H=22.3$ and $K=21.9$.
These images are available to the community, and have been
used by several groups presenting results at this symposium.
For further information and access to the data, please see
{\tt http://www.stsci.edu/ftp/science/hdf/clearinghouse/irim/irim\_hdf.html.}

A complete discussion of the infrared properties of galaxies in the HDF
is beyond the scope of this presentation.  Here I restrict my attention
to a few simple color properties of $z > 2$ Lyman break selected galaxies.
Although WFPC2 optical photometry in this paper is presented in
AB units, I will use standard (Vega--normalized) infrared magnitudes
in the discussion that follows.  For reference, the approximate conversion
to AB units for the $K_S$ bandpass used for the IRIM observations is 
$K_{AB} = K_{\rm Vega} + 1.86$.

\subsection{Colors and luminosities}

Figure~12 plots $V_{606} - K$ colors for galaxies in the HDF
with spectroscopically confirmed redshifts (plus a few stars).
At $0 < z \simlt 1$, galaxy colors are nicely bounded by the
range expected for star forming ``irregulars'' to old ellipticals.
The ``red envelope'' of galaxies at $z \approx 1$ is only slightly 
bluer than are giant elliptical galaxies in the local universe.
(Indeed, perhaps the most luminous galaxy in the HDF at rest--frame
optical wavelengths is a giant elliptical at $z=1.012$.)
Figure 13 shows shows the colors of nearly all 
optically selected \U300--dropout galaxies in the HDF brighter 
than $\V606 = 27$.  (A few have been excluded due to photometric
confusion with other objects in the near--IR data).  Most
Lyman break galaxies with $\V606 < 25.5$ are detected in 
the infrared images.  Beyond that magnitude there is increasing
incompleteness due to the limited depth of the IR data.

\begin{figure}
\centerline{\psfig{figure=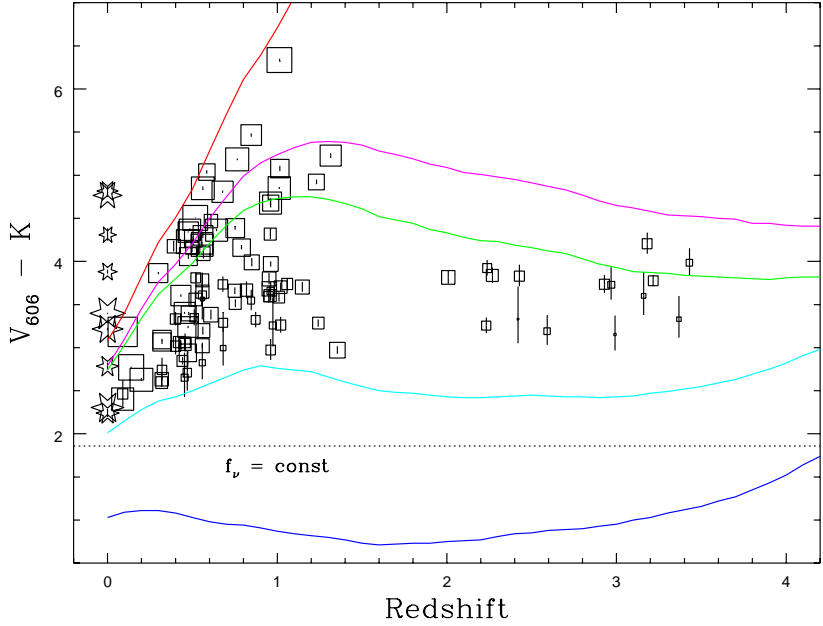,height=3.5in,angle=-90}}
\caption{Color versus redshift for galaxies with spectroscopic 
identifications in the HDF.  The symbol sizes scale inversely with
apparent $K$ magnitude;  star--shaped symbols at $z=0$ are known 
galactic stars.  The solid lines are {\it unevolving} tracks of 
$k$--corrected color vs.\ redshift for various model galaxy spectral
energy distributions designed to span the range (reddest to bluest)
from old ellipticals to actively star forming irregulars.  The bluest
model track, with $V_{606} - K \approx 1$ at most redshifts, is an 
extremely young (age = $10^7$ year) unreddened starburst.  
There are no spectroscopically observed galaxies in the HDF
which approach this color.}
\end{figure}

Lyman break galaxies mostly have colors in the range
$2.5 < \V606 - K < 4.5$ (although we cannot rule out 
that some of the fainter objects are intrinsically bluer).
As can be seen from the model tracks in figure~12 these colors
(shifted to the UV--optical rest frame)
would be normal for actively star forming spirals (Sb--Sc) in the local 
universe.   They are all much redder (as indeed are all HDF galaxies) 
than the expected colors of an extremely young, unreddened starburst
spectrum -- indeed, there are few faint galaxies anywhere in
the universe which are as blue or bluer than a flat ($f_\nu$)
spectrum over a long (optical--IR) wavelength baseline.   
The observed color does not, of course, tell us {\it why} 
the Lyman break galaxies have these colors; the effects 
of dust extinction on the rest--frame UV emission could 
have a significant impact on the observed colors (see previous 
section).  The typical (``$L^\ast$'' from figure 9) Lyman break
galaxy at $z \approx 3$ has $K = 21.5 \pm 1$, corresponding to
rest--frame $M_V = -22 \pm 1$ for $H_0 = 50$, $q_0 = 0.5$, 
i.e. similar to $L_V^\ast$ locally.

\begin{figure}
\centerline{\psfig{figure=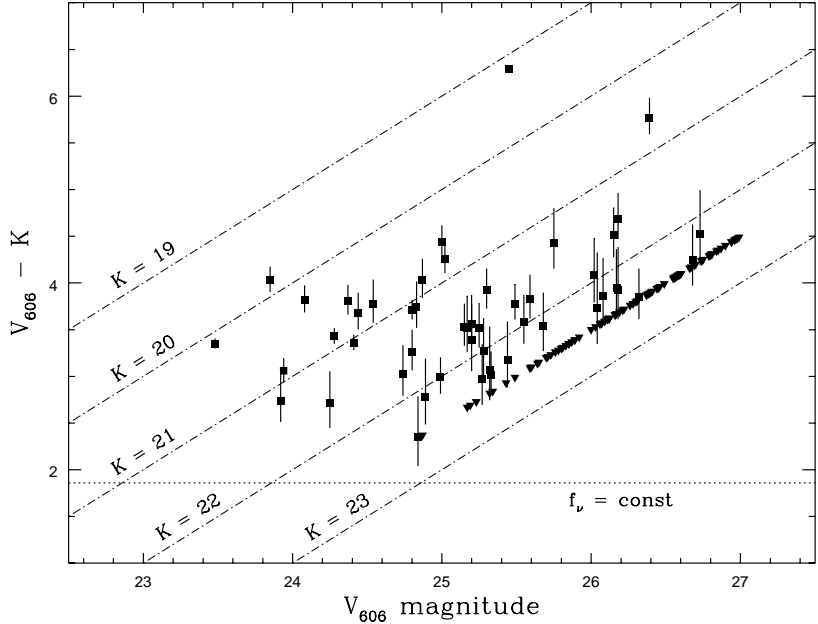,height=3.5in,angle=-90}}
\caption{$V_{606} - K$ colors of $U_{300}$ Lyman break selected
galaxies in the HDF.   Note that only a small subset of these galaxies
have spectroscopic redshifts.  Objects with $K > 22.5$ (i.e. beyond 
the robust detection limit of our IR data) are plotted as lower limits 
(triangles) at that magnitude.}
\end{figure}

\begin{figure}
\centerline{\psfig{figure=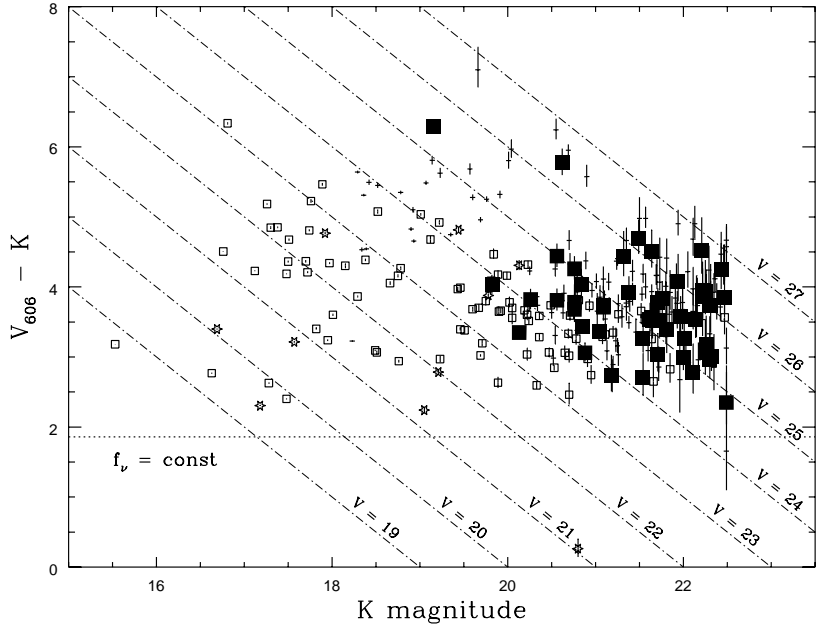,height=3.5in,angle=-90}}
\caption{$V_{606} - K$ color versus $K$ magnitude for objects with
$K < 22.5$ and $V_{606} < 27.0$ in the HDF.  Lyman break selected
galaxies are plotted as filled boxes.  Galaxies with spectroscopic
redshifts $0 < z < 2$ are plotted as unfilled boxes; galactic stars
are shown as star symbols.  Objects without spectroscopic observations
are plotted as small crosses.}
\end{figure}

Figure~14 shows a color--magnitude diagram for a $K$--selected 
sample of objects in the HDF.  Different symbol types code 
the objects as \U300--dropout Lyman break objects (with or 
without spectroscopic confirmation), galaxies with spectroscopically 
measured $z < 2$, stars, or unobserved objects (but which presumably 
mostly have $z < 2$, since they do not qualify as Lyman break 
candidates).  The optical--IR colors of the Lyman break galaxies are 
quite typical for the general population of faint galaxies in
the HDF.

Two galaxies in figures 13 and 14 have significantly redder
optical--IR colors ($\V606 - K$) than the majority of Lyman break 
objects.   Although they both satisfy the color selection criteria
defined in \S 2.1, it is not certain that these are also at $z > 2$.
In addition to their relatively bright infrared magnitudes, 
they are also much redder in $\V606 - \I814$ than the majority of
Lyman break galaxies ($\V606 - \I814 \approx 0.7$ to 1.0 -- compare
with figure 4), and their overall spectral energy distributions are
quite different than those of the known $z > 2$ galaxies in the HDF.
At present, neither object has a spectroscopic redshift;  future
observations should establish whether these are unusually red
and luminous members of the Lyman break population or interlopers
into the color selected sample from lower redshifts.

\subsection{Morphologies}

Existing ground--based near--IR images of Lyman break galaxies
in the HDF and elsewhere are mostly inadequate for answering questions
about the rest--frame optical morphologies of these objects.  
Our KPNO IRIM images of the HDF, for example, have a resolution
of $\approx 1\arcsec$, much larger than the typical scale 
lengths associated with the rest--frame UV light
seen in WFPC2 images (Giavalisco et al.\ 1996b).  However, 
a few simple conclusions can be drawn from the existing data.
In general, the centroid position of the infrared emission is 
spatially coincident with that of the optical images, 
suggesting that in most cases there is no substantial
``displacement'' between the dominant sources of rest--frame UV 
and optical luminosity in these objects.  In the HDF, however,
there is one spectacular exception found in galaxy 2--585.1, the
$z=2.01$ object whose redshift was serendipitously measured by
Elston et al.\ (see \S 2.2 above).   The WFPC2 images of this galaxy
show a chain--like assembly of blobs $\sim 3\arcsec$ long, making
it rather unusual among the Lyman break objects, which are mostly
quite compact.   The near--IR emission is spatially extended,
with its peak/centroid located just beyond one end of the ``chain,''
coincident with diffuse optical emission seen in the WFPC2 images.
In this object it appears that spatially segregated
stellar populations and/or extinction are affecting the observed
morphology -- the high surface brightness, star forming ``blobs''
of the chain structure protrude radially outward from a larger,
red host galaxy.   

\begin{figure}
\centerline{\psfig {figure=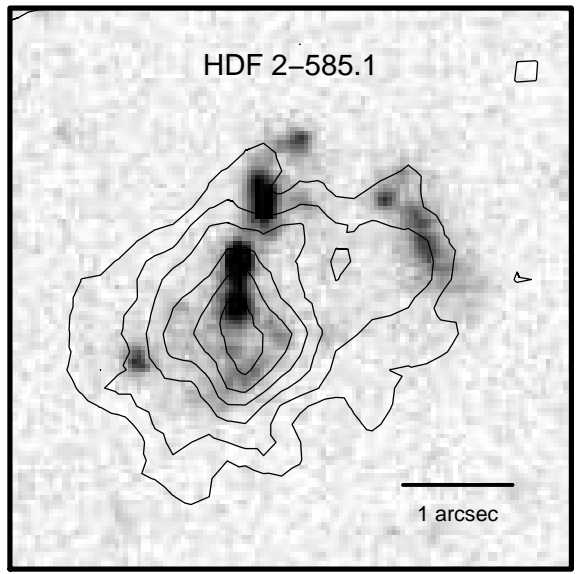,height=3in}}
\caption{$V_{606}$ WFPC2 image of the $z=2.01$ Lyman break
galaxy 2--585.1 (greyscale) overlaid with contours from the
ground--based infrared image.  Although the $\approx 1\arcsec$
resolution of the ground--based image does not allow morphological
details to be discerned, there is a clear spatial offset between
the peak of the IR emission and the brightest structures visible
in the rest--frame UV.  The object toward the upper right is not
a Lyman break candidate and thus is unlikely to be related to 
2--585.1.}
\end{figure}

\section{Summary} 

In this presentation I have tried to use HDF data to illustrate the 
ways in which multi--color photometry can be used first to select 
galaxies at high redshift and then to learn something about their 
intrinsic properties.   The great depth of the HDF and the precision 
of its photometry makes it ideal for illustrating and applying the 
Lyman break color selection technique:  cf.\ the remarkable prominence 
of the high redshift ``plume'' in figures 2 and 4.   
By \V606 = 26.5, nearly 1 in 4 HDF galaxies 
is a Lyman break candidate and thus is likely to be at $z > 2$.
Spectroscopy of HDF galaxies at all redshifts has proceeded at 
a remarkable pace.  After only two observing seasons, the central 
HDF is almost certainly the piece of celestial real estate with
the highest surface density of measured galaxy redshifts 
(now $\approx 24$/arcmin$^{2}$, of which $\sim$20\% are at $z > 2$).

The distribution of ultraviolet luminosities of $z \approx 3$
galaxies, converted to star formation rates using a simple prescription,
spans a similar range to that of galaxies in the local universe.  
Schechter function fits give a characteristic 1500\AA\ specific 
luminosity of $M_{AB} \approx -21$ (for $H_0 = 50$, $q_0 = 0.5$), 
corresponding to a star formation rate of $\sim 14 M_\odot$/year.
Very few Lyman break galaxies have ``raw'' SFRs $> 50 M_\odot$/year.
If these measurements are taken at face value, these objects cannot 
produce the total stellar mass of $\sim L^\ast$ galaxies in short 
timescales as traditional ``monolithic'' formation scenarios for 
elliptical galaxies would require.  This, then, would suggest
that massive galaxy formation takes place either at still higher
redshifts where we have yet to look, or proceeds by the hierarchical
assembly of smaller objects as expected in theories such as CDM.
However, the possible extinction corrections to the UV luminosities of 
Lyman break galaxies are highly uncertain and could be quite large.  
Their UV colors are redder than expected from spectral models of
``naked'' star forming galaxies, a fact which could easily be
explained by the presence of dust.  The derived extinction 
corrections based on these colors, however, are extremely sensitive 
to the form of the extinction law in the ultraviolet, and can range 
from factors of 2 to $> 7$.  New observations in the infrared, and 
eventually at far--IR and sub--millimeter wavelengths, will provide 
an independent measure of star formation rates which can be useful 
for addressing this question.

Characteristic rest--frame optical luminosities of Lyman break
galaxies, as measured from infrared photometry, are $M_V = -22$.  
Their UV--optical rest--frame colors galaxies span a range which would be 
typical for normal spirals in the nearby universe.  Detailed morphological 
study of $z \approx 3$ galaxies at rest--frame optical wavelengths must 
await observations with NICMOS, which will take place in 1997--1998.

Although Lyman break color selection is a simple technique compared 
to more sophisticated photometric redshift methods, it has the virtue 
of being relatively model independent and easy to apply and understand.
Like all such methods, however, its utility depends strongly on the 
degree to which it is tested and calibrated by follow--up spectroscopy, 
which requires substantial effort on large ground--based telescopes.   
With hundreds of Lyman break redshifts in hand, we can begin to carry 
out fairly sophisticated analyses of the luminosity function, clustering, 
and other properties of galaxies at $z \approx 3$.  For the HDF, despite 
the impressive observing efforts to date, we are unlikely ever to succeed 
in collecting hundreds of redshifts for $z > 2$ galaxies with existing 
telescopes and instrumentation.  However, we may take advantage of the 
insights gained from studying the Lyman break galaxy population in 
non--HDF data sets to aid interpretation of the HDF objects, and thus to 
use the HDF to push the method to different flux and redshift limits.  

Regardless of how much we learn about high redshift galaxies in the 
HDF, we must remind ourselves of what a small volume the HDF probes.  
The entire comoving volume over which \U300--dropouts have been
found in the HDF, $2 < z < 3.5$, is only 18000$h_{50}^{-3}$~Mpc$^3$
for $q_0 = 0.5$.   Locally, that would correspond to a sphere with 
radius 16.2$h_{50}^{-1}$ Mpc -- not even reaching the Virgo cluster!
We must therefore be cautious about how representative HDF galaxies
are in any statistical sense, particularly in light of recent evidence
for strong clustering at $z > 2$.    The pencil--beam geometry
of the HDF volume ensures that it will traverse a wider range of 
large scale structures than would the corresponding 14.2~Mpc radius 
sphere locally, but such a geometry brings its own complications
for some applications.  For example, clustering may introduce large 
fluctuations in $N(z)$ which can seriously complicate analyses of the 
redshift evolution of galaxy properties, global luminosity density, etc.
It is therefore risky to extrapolate too far from the HDF to the 
properties of galaxies in the high redshift universe as a whole.  
Ultimately, however, the insights gathered from the HDF, when 
calibrated with data from ground--based, large--volume surveys, 
should provide a powerful means of understanding the early stages 
in the evolution of normal galaxies.
 
\begin{acknowledgments}

I would like to extend special thanks to my collaborators 
Chuck Steidel, Mauro Giavalisco,
Max Pettini, Kurt Adelberger, and Mindy Kellogg, for endlessly
interesting discussions, much hard work, and for allowing me to present 
materials in advance of publication.  The same thanks also go to 
Adam Stanford, Peter Eisenhardt, Richard Elston, and Matt Bershady 
for their collaboration on the KPNO infrared imaging program.  
Finally, I would also like to thank my colleagues at STScI from 
the HDF Team, and the editors of this volume for their considerable
patience.

\end{acknowledgments}

% \begin{thebibliography}{} 
\vspace{1cm}
\centerline{REFERENCES}
\vspace{0.5cm}

\reference    Calzetti, D., Kinney, A.L., \& Storchi--Bergmann, T.,
	1994, \apj, 429, 582.

\reference    Calzetti, D., 1997, in {\it The Ultraviolet Universe at Low 
	and High Redshift,} ed. W. Waller (Woodbury: AIP Press), in press 
	(astro-ph/9706121).

\reference   Cohen, J.G., Cowie, L.L., Hogg, D.W., Songaila, A.,
	Blandford, R., Hu, E.M., \& Shopbell, P., 1996, \apj, 471, L5.

\reference   Colley, W.M., Rhoads, J.E., Ostriker, J.P., \& Spergel, D.N.,
	1996, \apj, 473, L63.

\reference   Gallego, J., Zamorano, J., Arag\'on--Salamanca, A., \& Rego, M.,
	1995, \apj, 455, L1.

\reference   Franx, M., Illingworth, G., Kelson, D.D., Van Dokkum, P.G.,
	\& Tran, K., 1997, \apj, 486, L75.

\reference   Giavalisco, M., Livio, M., Bohlin, R.C., Macchetto, F.D.,
	\& Stecher, T.P., 1996a, \aj, 112, 369.

\reference   Giavalisco, M., Steidel, C.C., \& Macchetto, F.D., 1996b, 
	\apj, 470, 189.

\reference   Giavalisco, M., Steidel, C.C., Adelberger, K.L., Dickinson, M., 
	Pettini, M., \& Kellogg, M., 1998, \apj, submitted.

\reference   Guhathakurta, P., Tyson, J.A., \& Majewski, S.R., 1990, 
	\apj, 357, L9.

\reference   Hogg, D.W., Blandford, R., Kundic, T., Fassnacht, C.D.,
	\& Malhotra, S., 1996, \apj, 467, L73.

\reference   Hogg, D.W., Neugebauer, G., Armus, L., Matthews, K.,
	Pahre, M.A., Soifer, B.T., \& Weinberger, A.J., 1997, \aj, 113, 2338.

\reference   Kennicutt, R.C., 1983, \apj, 272, 54.

\reference   Lowenthal, J.D., Koo, D.C., Guzman, R., Gallego, J., 
	Phillips, A.C., Faber, S.M., Vogt, N.P., Illingworth, G.D.,
	\& Gronwall, C., 1997, \apj, 481, 673.

\reference   Madau, P., 1995, \apj, 441, 18.

\reference   Madau, P., Ferguson, H.C., Dickinson, M., Giavalisco, M., 
	Steidel, C.C., \& Fruchter, A., 1996, \mnras, 283, 1388.

\reference   Madau, P., Pozzetti, L, \& Dickinson, M., 1998, 
	\apj, in press (astro-ph/9708220).

\reference   Meurer, G.R., Heckman, T.M., Lehnert, M.D., Leitherer, C.,
	\& Lowenthal, J., 1997, \aj, 114, 54.

\reference    Pettini, M., Steidel, C.C., Dickinson, M., Kellogg, M.,
	Giavalisco, M., \& Adelberger, K.L., 1997, in 
	{\it The Ultraviolet Universe at Low and High Redshift,} 
	ed. W. Waller (Woodbury: AIP Press), in press (astro-ph/9707200).

\reference   Songaila, A., Cowie, L.L., \& Lilly, S.J., 1990, \apj, 348, 371.

\reference   Steidel, C.C., \& Hamilton, D., 1992, \aj, 104, 941.

\reference   Steidel, C.C., Pettini, M., \& Hamilton, D., \aj, 110, 2519.

\reference   Steidel, C.C., Giavalisco, M., Pettini, M., Dickinson, M.,
	\& Adelberger, K.L., 1996a, \apj, 462, L17.

\reference   Steidel, C.C., Giavalisco, M., Dickinson, M., \& Adelberger, K.L.,
	1996b, \aj, 112, 352.

\reference   Steidel, C.C., Adelberger, K.L., Dickinson, M., Giavalisco, M.,
	Pettini, M., \& Kellogg, M., 1998, \apj, 492, 428.

\reference   Tresse, L., \& Maddox, S.J., 1997, \apj, in press 
	(astro-ph/9709240).

\reference   Treyer, M.A., Ellis, R.S., Milliard, B., \& Donas, J.,
	1997, in {\it The Ultraviolet Universe at Low and High Redshift,}
	ed. W. Waller (Woodbury: AIP Press), in press (astro-ph/9706223).

\reference   Warren, S.J., Hewitt, P.C., Irwin, M.J., McMahon, R.G., 
	Bridgeland, M.T., Bunclark, P.S., \& Kibblewhite, E.J., 1987, 
	Nature, 325, 131.

\reference   Williams, R.E., et al.\ 1996, \aj, 112, 1335.

\reference   Zepf, S.E., Moustakas, L.A., \& Davis, M., 1997, \apj, 474, L1.

% \end{thebibliography}

\end{document}